\documentclass[11pt]{article}
\usepackage{cospar}
\usepackage{url}

\usepackage{graphicx}
\usepackage[figuresright]{rotating}

\title{COMET AND ASTEROID HAZARD TO THE TERRESTRIAL PLANETS}

\author{S. I. Ipatov\address{(1) NRC/NAS senior research associate, 
NASA/GSFC, Mail Code 685,
Greenbelt, MD 20771, USA (current address);
(2) Institute of Applied Mathematics,
Miusskaya sq. 4, Moscow 125047, Russia} and
J. C. Mather\address{NASA/GSFC, Mail Code 685, Greenbelt,
MD 20771, USA}}

\begin{document}

\maketitle

\begin{abstract}

We estimated the rate of comet and asteroid collisions with the 
terrestrial planets by calculating the orbits of 13000 
Jupiter-crossing objects (JCOs) and 1300 resonant asteroids and 
computing the probabilities of collisions based on random-phase 
approximations
and the orbital elements sampled with a 500 yr step. 
The Bulirsh-Stoer and a symplectic orbit integrator 
gave similar results
for orbital evolution, but may give different collision
probabilities with the Sun.   
A small fraction of former JCOs reached 
orbits with aphelia inside Jupiter's orbit, and some reached  Apollo 
orbits with semi-major axes less than 2 AU, Aten orbits, and 
inner-Earth orbits (with aphelia less than 0.983 AU) and remained 
there for millions of years. Though less than 0.1\% of the total, 
these objects were responsible for most of the collision probability 
of former JCOs with Earth and Venus. 
We conclude that a significant 
fraction of near-Earth objects could be extinct comets that came from 
the trans-Neptunian region.

\end{abstract}

\section*{INTRODUCTION}

The main asteroid belt, the trans-Neptunian belt, and the Oort cloud 
are considered to be the main sources of the objects that could 
collide with the Earth. 
Reviews of the asteroid and comet hazard were given by Ipatov (2000, 
2001), and Bottke et al. (2002). Many scientists, e.g. Bottke et al. 
(2002), Binzel et al. (2002), and Weissman et al. (2002), 
believe that asteroids are the main source of near-Earth 
objects (NEOs, i.e. objects with perihelion distance $q$$<$1.3 AU).
Bottke et al. (2002) considered that
there are 200$\pm$140 km-sized 
Jupiter-family comets 
at $q$$<$1.3 AU, with $\sim$80\% of them being extinct comets.
Duncan et al. (1995) and Kuchner (2002) investigated the migration of 
trans-Neptunian objects (TNOs)
to Neptune's orbit, and  Levison and Duncan (1997) studied the 
migration from Neptune's orbit to Jupiter's orbit. Ipatov and Hahn 
(1999) considered the migration of 
Jupiter-crossing objects (JCOs) 
with initial orbits
close to the orbit of Comet P/1996 R2 and found that on average such 
objects spend 
about 5000 yr
in orbits which cross both the orbits of Jupiter and Earth. Using 
these results and additional orbit integrations, and assuming that 
there are  $5\times10^9$ 1-km TNOs with 30$<$$a$$<$50 AU (Jewitt and 
Fernandez, 2001), Ipatov (2000, 2001) found that about $10^4$ 1-km 
former TNOs are Jupiter-crossers now and  10-20\% or more 1-km 
Earth-crossers could have come from the Edgeworth-Kuiper belt into 
Jupiter-crossing orbits. 
In the present paper we use the estimates by Ipatov (2001), 
but now include a much  larger number of JCOs. Preliminary results 
were presented by  Ipatov (2002, 2003), who also discussed the 
formation of TNOs and asteroids.

\section*{PROBABILITIES OF COLLISIONS OF NEAR-EARTH OBJECTS WITH 
PLANETS IN THE MODEL OF FIXED ORBITAL ELEMENTS}

As the actual collisions of migrating objects with terrestrial 
planets are rare, we use an approximation of random phases and 
orientations to estimate probabilities of collision for families of 
objects with similar orbital elements. We suppose that their 
semi-major axes $a$, eccentricities $e$ and inclinations $i$ are 
fixed, but the orientations of the orbits can vary. 
When the orbit of a minor body crosses the orbit of  
a planet at a distance
$R$ from the Sun, the characteristic time to collide, $T_f$, is a 
factor of $k =v/v_c = \sqrt{2a/R -1}$ times that computed with an 
approximation of constant velocity, where $v$ is the velocity at the 
point where the orbit of the body crosses the orbit of the planet, 
and $v_c$ is the velocity for the same semi-major axis and a circular 
orbit.  This coefficient $k$  modifies the formulas obtained by 
Ipatov (1988a, 2000) for characteristic collision and close encounter 
times of two objects moving around the Sun in crossing orbits. These 
formulas 
depend also on the synodic period and improve on 
\"Opik's formulas when the semi-major axes of the  objects are close 
to each other. As an example, at $e$=0.7 and $a$=3.06 AU,  we have 
$k$=2.26.

Based on these formulas, we calculated probabilities (1/$T_f$) 
for $\sim$1300 
NEOs, including  343 Venus-crossers, 756 Earth-crossers and 1197 
Mars-crossers. The  values of $T_f$ (in Myr), $k$, and the number 
$N_f$ of  objects considered are presented in Table 1. We considered 
separately the Atens, Apollos, Amors, and several Jupiter-family 
comets (JFCs). The relatively small values of $T_f$ for Atens and for 
all NEOs colliding with the Earth are due to several Atens with small 
inclinations discovered during the last three years. If we increase 
the inclination of the Aten object 2000 SG344 from $i$=$0.1^\circ$ to 
$i$=$1^\circ$, then for collisions with the Earth we find $T_f$=28 Myr 
and $k$=0.84 for Atens and $T_f$=97 Myr and $k$=1.09 for NEOs.  These 
times are much longer, and illustrate the importance of rare objects. 
Due to observational biases actual values of $T_f$ can be greater 
than those in Table 1.  


\begin{table}[h]
\begin{center}
\begin{minipage}{14.3cm}

\caption{Characteristic collision times $T_f$ (in Myr)
of minor bodies with planets, coefficient $k$,
and number $N_f$ of simulated objects for the set of NEOs known in 2001.}
$
\begin{array}{lccccc}
\hline
       & $Atens$     & $Apollos$    & $Amors$ &  $NEOs$ & $JFCs$ \\
\hline
$Planet$&T_f~~k~~N_f& T_f~~k~~N_f& T_f~~k~~N_f& T_f~~k~~N_f& T_f~~k\\
$Venus$ & 106~~ 1.2 ~~ 94 & 186~~ 1.7 ~~ 248& -  & 154~~ 1.5 ~~ 343& 
2900~~ 2.5 \\

$Earth$ &15~~ 0.9 ~~110  & 164~~ 1.4 ~~643& 211~~ 2.0~~1 & 67~~ 
1.1~756  &2200~~ 2.3\\

$Mars$  &475~~ 0.4~~6  & 4250~~ 0.9~~574 &5810~~ 1.1~~616 &4710~~ 1.0 
~1197& 17000~~ 1.8 \\
\hline

\end{array}
$
\end{minipage}
\end{center}
\end{table}

\section*{ORBITAL EVOLUTION OF JUPITER-FAMILY COMETS AND RESONANT ASTEROIDS }

As the next step in estimating probabilities, we calculated the 
orbital evolution 
for thousands of test particles with 
initial
orbits similar to known 
comets and asteroids, but having slightly different initial 
conditions.  The results confirm that most of the collision 
probability comes from a handful of very rare cases in which the test 
particle 
is Earth-crossing for an 
extended period of time.  

For initial 
investigations of the migration of bodies under the gravitational 
influence of the planets, we used the integration package of Levison 
and Duncan (1994).  In most cases we omitted the influence of Mercury 
and Pluto. Here and in Tables 2-3 
and Figs. 1, 2 and 4 
we present the 
results obtained by
the Bulirsh-Stoer method (BULSTO code) with the integration step 
error 
less than
$\varepsilon$$\in$[$10^{-9}$-$10^{-8}$], and in the next 
section we compare them with those of BULSTO at 
$\varepsilon$$\le$$10^{-12}$ and  a symplectic method. 

\begin{table}[h]
\begin{center}
\begin{minipage}{18cm}

\caption{
Mean probability $P$$=$$10^{-6}P_r$ of a collision of an object with 
a planet (Venus=V, Earth=E, Mars=M)
during its lifetime,  mean time $T$ (in Kyr) during which 
$q$$<$$a_{pl}$, $T_c$=$T/P$ (in Gyr),
mean time $T_J$ (in Kyr) spent in Jupiter-crossing orbits, 
mean time $T_d$ (in Kyr) spent in orbits with $Q$$<$4.2 AU,
and ratio 
$r$ of times spent in Apollo and Amor orbits. 
Results from  BULSTO 
code at $\varepsilon$$\sim$$10^{-9}$-$10^{-8}$. $\Sigma$ denotes the sum for
several series presented in the above lines. 
For $N$=7349, 2P runs were excluded.}

$ \begin{array}{lllllcccccccccccl}

\hline

   & & &&&$V$ & $V$ & $V$ & $E$ & $E$ & $E$ & $M$ & $M$ & $M$ & & &\\

\cline{6-17}

  & N& a& e&i&P_r & T &T_c& P_r & T &T_c& P_r & T &T_c& r & T_J &T_d \\

\hline

n1&1900& &&& 2.42 & 4.23 &1.75& 4.51 & 7.94 &1.76&  6.15 & 30.0 
&4.88& 0.7 & 119& 20 \\
$2P$& 501 &2.22 &0.85&12^\circ&141 &345 &2.45 & 110 & 397 &3.61&10.5& 
430 &41.0& 18.& 173&249 \\
$9P$& 800&3.12 &0.52&10^\circ &1.34 &1.76 &1.31&3.72 & 4.11 
&1.10&0.71 & 9.73&13.7& 1.2 &96& 2.6 \\
$10P$& 2149&3.10&0.53&12^\circ &28.3 & 41.3&1.46& 35.6 & 
71.0&1.99&10.3 & 169.&16.4&1.6 &122&107\\
$22P$& 1000&3.47 &0.54&4.7^\circ &1.44&2.98&2.07&1.76&4.87&2.77&0.74 
& 11.0&14.9& 1.6 &116&1.5 \\
$28P$& 750 &6.91&0.78&14^\circ & 1.7 & 21.8&12.8& 1.9 & 34.7&18.3& 
0.44 & 68.9&157&1.9 &443&0.1 \\
$39P$& 750 &7.25&0.25&1.9^\circ & 1.06&1.72&1.62&1.19& 3.03&2.55& 
0.31 & 6.82&22 & 1.6 &94&2.7 \\
\Sigma & 7349&&&& 9.52 & 16.2&1.70 & 12.6 & 27.9&2.21 &  4.89 & 
62.4&12.8 & 1.4 &112&41 \\
\Sigma & 7850&&&& 17.9 & 37.7&2.11 & 18.8 & 51.5&2.74 &  5.29 & 
85.9&16.2 & 2.6 &116&54 \\
\Sigma & 7852&&&& 130. & 72.6 & 0.56 & 84.5 & 95.7 & 1.13 & 13.5 & 
132&9.81 & 3.8 &116&101 \\
$R2$& 24 & 3.79&0.31&2.6^\circ&0.53&0.6&1.13&2.84&1.6&0.56&6.97&14&2.0&&&\\
3:1& 288 & 2.5&0.15&10^\circ &1286 &1886&1.47& 1889& 2747 &1.45& 
488&4173&8.55 & 2.7 &229&5167 \\
5:2& 288 & 2.82&0.15&10^\circ &101   &173&1.71 & 318 & 371 &1.16&209 
&1455&6.96 & 0.5 &233& 1634 \\
\hline

\end{array} $
\end{minipage}
\end{center}
\end{table}

      In the first series of runs (denoted as $n1$) we calculated the 
evolution of 1900 JCOs moving in initial orbits close to those of 20 
real JCOs with period $5$$<$$P_a$$<$9 yr. In other series of runs, initial 
orbits were close to those of a single comet (2P, 9P, 10P, 22P, 28P, 
or 39P). For the 2P runs, we included Mercury in the integrations. We 
also investigated the orbital evolution of asteroids initially moving 
in the 3:1 and 5:2 resonances with Jupiter. For the JCOs we varied 
only the initial mean anomaly $\nu$.  The number of objects in one 
run usually was $\le$250. In most JCO cases  the time $\tau$ when 
perihelion was passed was varied with a step $d\tau$$\le$1 day
(i.e., $\nu$ was varied with a step $<$$0.2^\circ$). Near the $\tau$ 
estimated from observations, we used smaller steps. In most JCO cases 
the range of initial values of $\tau$ was less than several tens of 
days. For asteroids, we varied initial values of $\nu$ and the 
longitude of the ascending node from 0 to 360$^\circ$.
The approximate values of initial orbital elements
are presented in Table 2.
We initially integrated the orbits for  $T_S$$\ge$10 Myr. After 10 
Myr we tested whether some of remaining objects could reach inside 
Jupiter's orbit; if so,  the
calculations were usually continued. Therefore the results for orbits 
crossing or inside Jupiter's orbit were the same as if the 
integrations had been carried to 
lifetimes of objects. For Comet 2P and 
resonant asteroids, we integrated  until all objects were ejected 
into hyperbolic orbits or collided with the Sun. In some previous 
publications we have used smaller $T_S$, so these new data are more 
accurate.

      In our runs, planets were considered as material points so 
literal collisions did not occur.  However, using the formulas of the 
previous section, and the orbital elements sampled with a 500 yr 
step, we calculated the mean probability  $P$ of collisions. We 
define $P$ as  $P_\Sigma/N$, where $P_\Sigma$ is the probability for 
all $N$ objects of a
collision of an object with a planet during its lifetime,   the mean 
time $T$=$T_\Sigma/N$ during which
perihelion distance $q$ of an object was less than the semi-major 
axis $a_{pl}$ of the planet,
and the mean time $T_J$ during which an object moved in 
Jupiter-crossing orbits. The values of $P_r$=$10^6P$,
$T_J$ and $T$ are shown in Table 2. Here $r$ is the ratio of the 
total time interval when orbits are of Apollo
type ($a$$>$1 AU, $q$=$a(1-e)$$<$$1.017$ AU) at $e$$<$0.999 to that 
of Amor type ($1.017$$<$$q$$<$1.3 AU) and $T_c$=$T/P$ (in Gyr).
In almost all runs $T$ was equal to the mean time 
in orbits which cross the orbit of the planet
and 1/$T_c$ was a probability
of a collision per year (similar to 1/$T_f$).
      The  results showed that most of the probability of collisions of
considered former JCOs with the terrestrial planets is due to a 
few objects each of which
orbited for several Myr with aphelion 
$Q$$<$4.2 AU.
Some had typical asteroidal and NEO orbits and reached $Q$$<$3 AU for 
several Myr. 
Large
values of    
$P$ for Mars in the $n1$ runs were caused by a single object with a 
lifetime of 26 Myr. 
Total times spent by $N$ JCOs and asteroids during their lifetimes 
in orbits typical for
inner-Earth objects (IEOs, $Q$$<$0.983 AU, Michel et al., 2000), Aten 
($a$$<$1 AU and
$Q$$>$0.983 AU), Al2 ($q$$<$1.017 AU and 1$<$$a$$<$2 AU), Apollo, and 
Amor objects
are presented in Table 3.
These 
times for Earth-crossing objects were mainly due to 
a few tens of objects
   with high collision probabilities.
Of
the JCOs with initial orbits close to those of 
10P and 2P, six and nine respectively moved into Apollo orbits with 
$a$$<$2 AU (Al2 orbits) for at least 0.5 Myr each, and five of them 
remained in such orbits for more than 5 Myr each. The contribution of 
all the other objects to Al2 orbits was smaller. 
Only one and two JCOs reached IEO and Aten orbits, respectively.

%
\begin{table}[h]

\begin{center}
\begin{minipage}{13.4cm}
\caption{
Times (in Myr) spent by $N$ JCOs and 
asteroids during their lifetimes, with 
number of such objects in [ ]. Results from BULSTO 
code at $\varepsilon$$\sim$$10^{-9}$-$10^{-8}$.
}
$ \begin{array}{lllllllll}

\hline

   & N& $IEOs$& $Aten$&  $Al2$& $Apollo$ & $Amor$ & a>5 $ AU$ \\

\hline
$JCOs$ & 7852 & 10 $ $[1] & 86 $ $ [2] & 411 $ $ [43] & 659 & 171 & 7100 \\

$JCOs without 2P$ & 7350 & 10 $ $ [1]& 3.45 $ $ [1] & 23 $ $[10]& 207 & 145 & 7000 \\
3:1 $ resonance$& 288   & 13 $ $ [2] & 4.5 $ $[4] & 433 $ $[27]$ $ & 790 $ $ & 290 $ $ 
& 83 $ $ \\

5:2 $ resonance$& 288 & 0 & 0 & 17 $ $[5] & 113 $ $ & 211 $ $ & 253 $ $\\

\hline

\end{array} $
\end{minipage}
\end{center}

\end{table}

One former JCO (Fig. 1a), which had an initial orbit close to that of 10P, 
moved in Aten orbits for 3.45
Myr, and the probability of its collision with the Earth from such 
orbits  was 0.344 (so $T_c$=10 Myr was even smaller than the values of $T_f$
presented in Table 1; i.e., this object had smaller $e$ and $i$
than typical observed Atens),  greater
than that  for the 
7850 other simulated former JCOs during 
their lifetimes (0.15). It also moved for
about 10 Myr  in IEO 
orbits before its collision with Venus, 
and during this time the
probability $P_V$=0.655 of its collision with Venus was greater 
($P_V$$\approx$3 for the time interval presented in Fig. 1a) 
than 
that  for the 7850 JCOs during their lifetimes (0.14).
At $t$=0.12 Myr orbital elements of this object jumped considerably
and the Tisserand parameter increased from $J$$<$3 to $J$$>$6, and $J$$>$10
during most of its lifetime.
Another object (Fig. 1b) moved in highly eccentric Aten orbits for 83 Myr, and 
its lifetime before collision
with the Sun was 352 Myr. Its probability of collisions with Earth, 
Venus and Mars during its
lifetime was 0.172, 0.224, and 0.065, respectively. These two objects 
were not included in Table 2 except for the entry for $N$=7852.
The data for Comet P/1996 R2 (line R2) were not included in the sums.
Ipatov (1995) obtained the migration of JCOs into IEO and Aten 
orbits using the approximate
method of spheres of action for taking into account the gravitational 
interactions of bodies with planets.
The mean time $T_E$ during which a JCO was moving in Earth-crossing 
orbits is $9.6\times10^4$ yr for
the 7852 simulated JCOs, and $\approx8\times10^3$ yr for the $n1$ case.
The ratio $P_S$ of the number of objects colliding with the Sun to 
the total number of
escaped (collided or ejected) objects was less than 0.015 
for the considered runs (except for 2P).


Some former JCOs 
spent a long time in the 3:1 resonance with Jupiter
and with 2$<$$a$$<$2.6 AU. Other objects reached Mars-crossing orbits 
for long times. We conclude that JCOs can supply bodies to the 
regions which are considered by many scientists (Bottke et al., 2002) 
to belong to the main sources of NEOs.

\begin{figure}

\includegraphics[width=60mm]{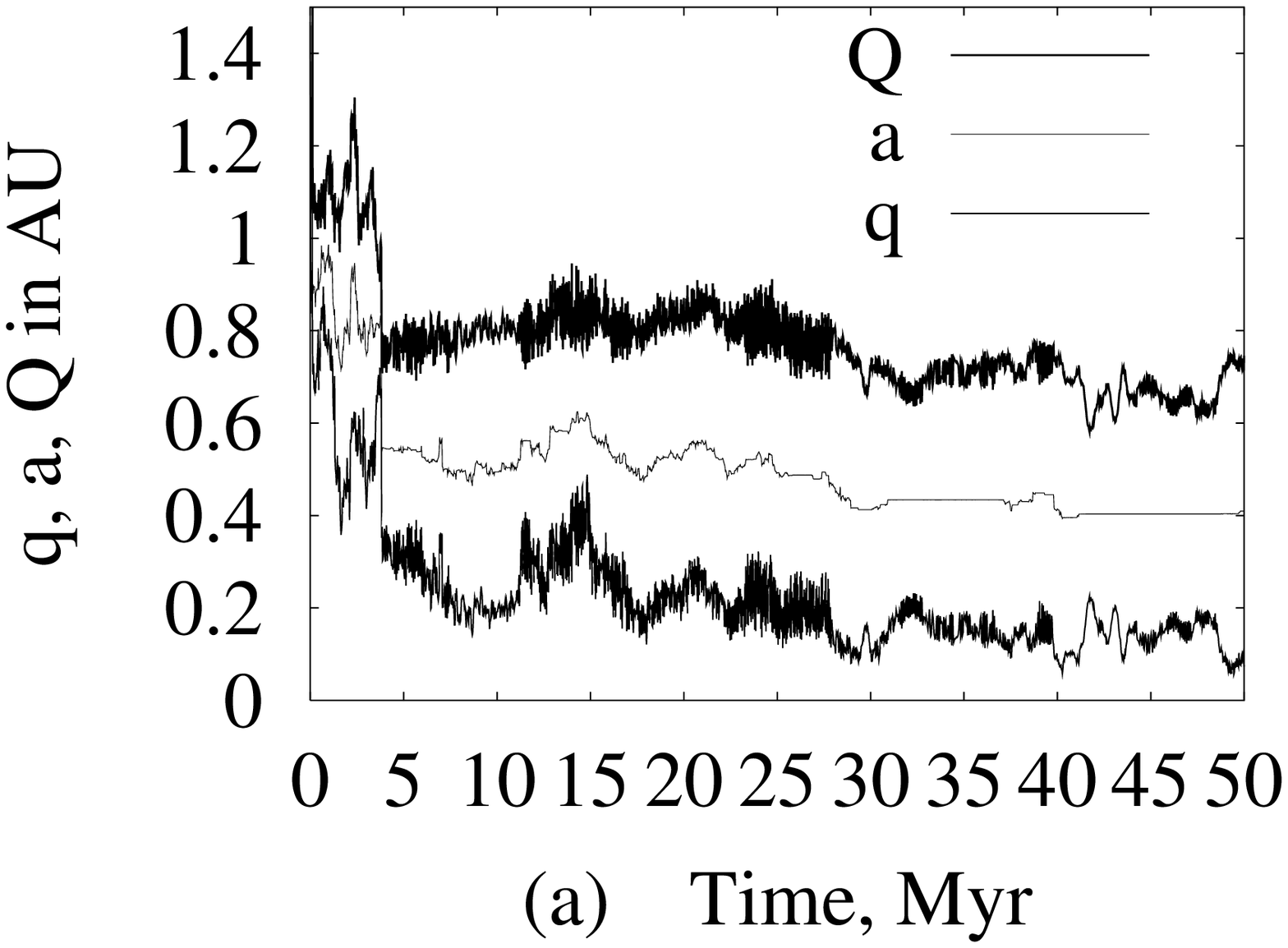}
\includegraphics[width=60mm]{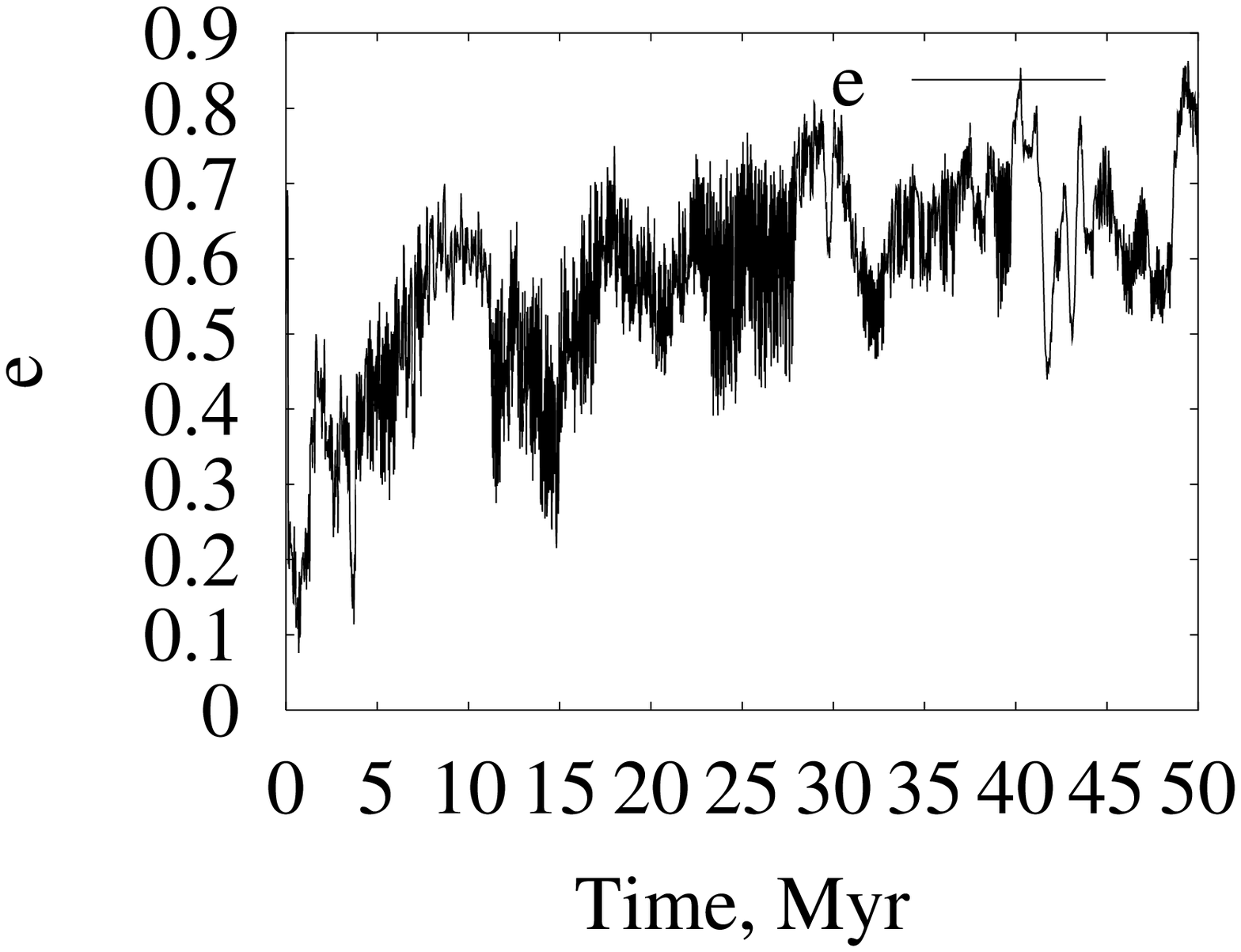}
\includegraphics[width=60mm]{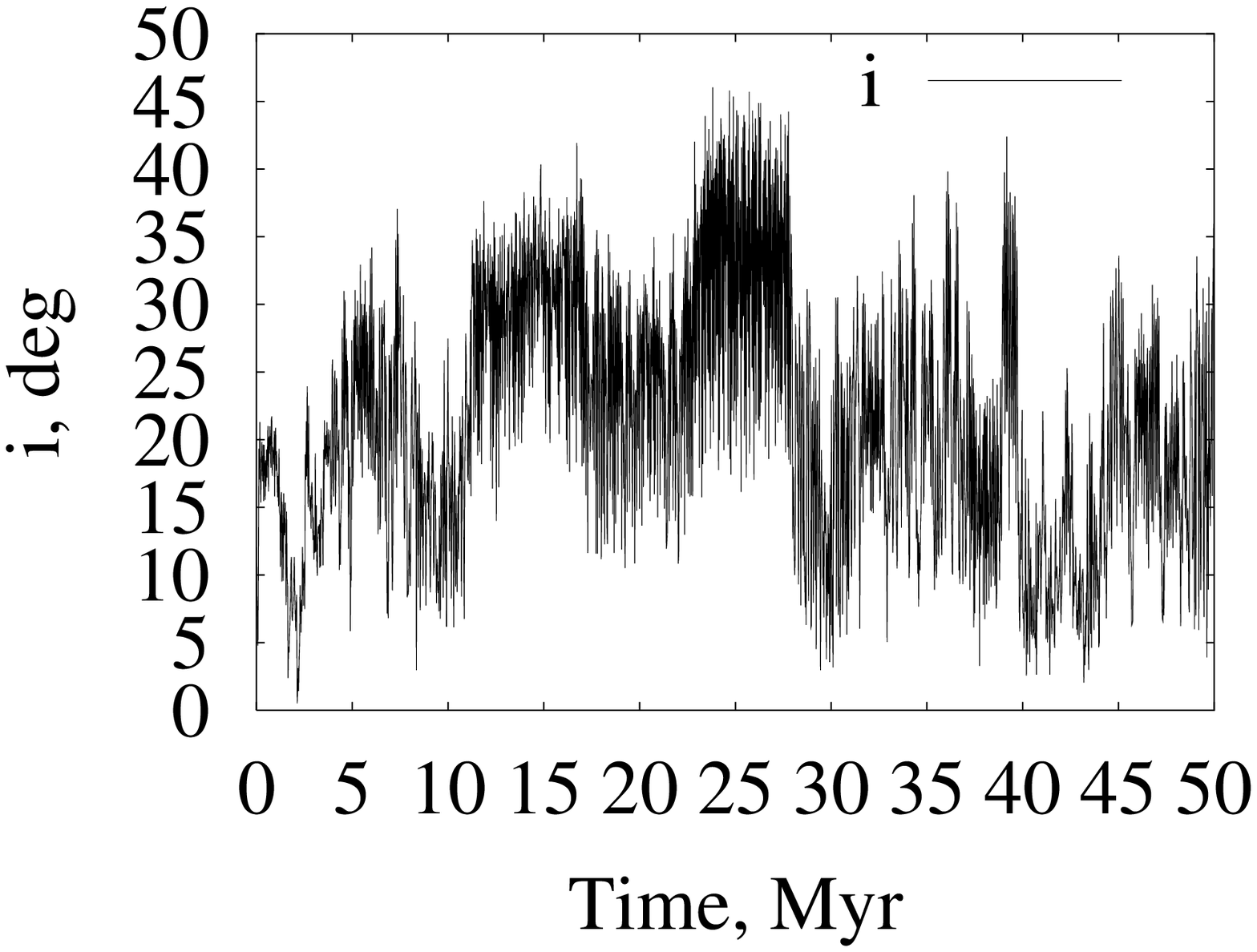}

\includegraphics[width=60mm]{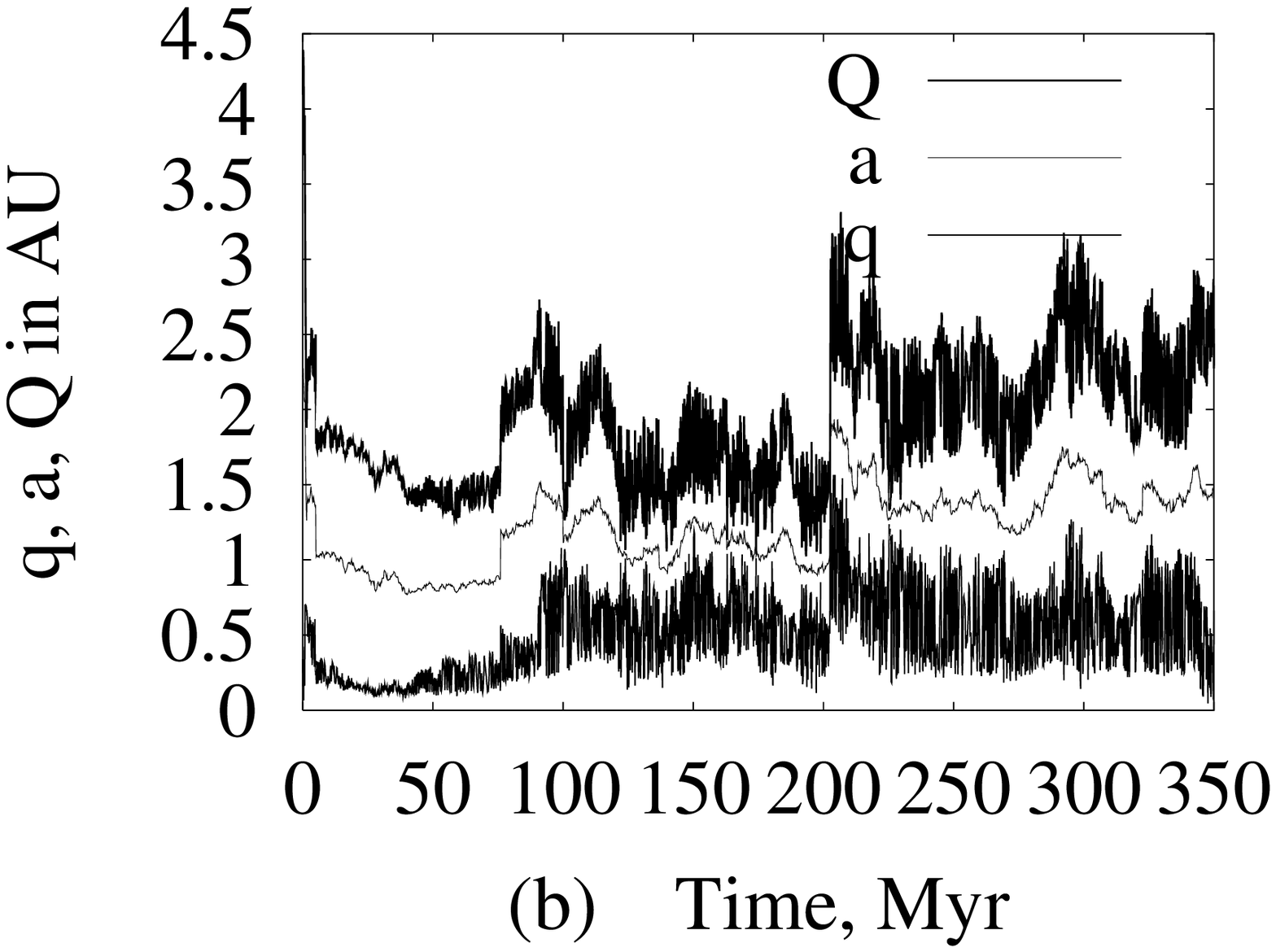}
\includegraphics[width=60mm]{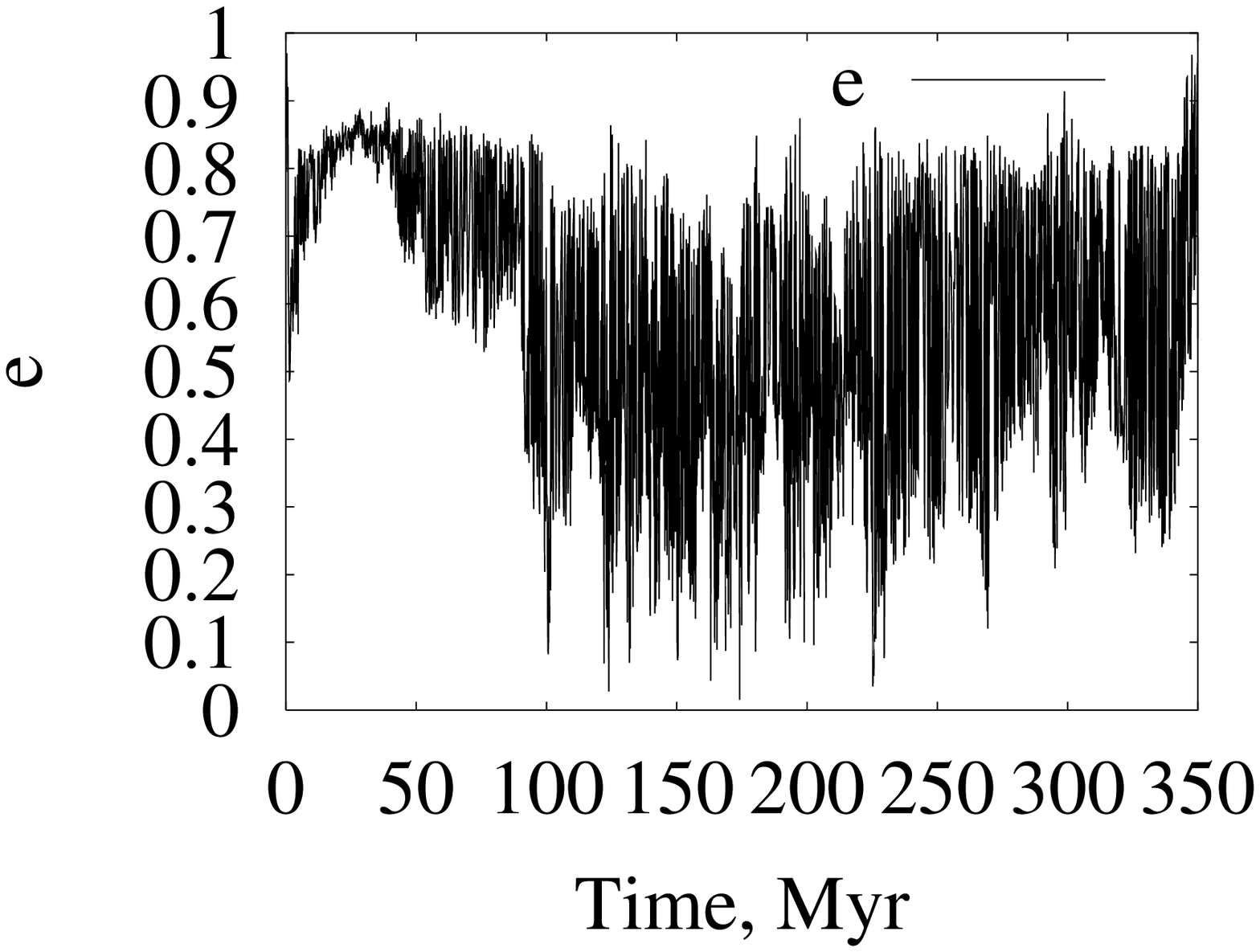}
\includegraphics[width=60mm]{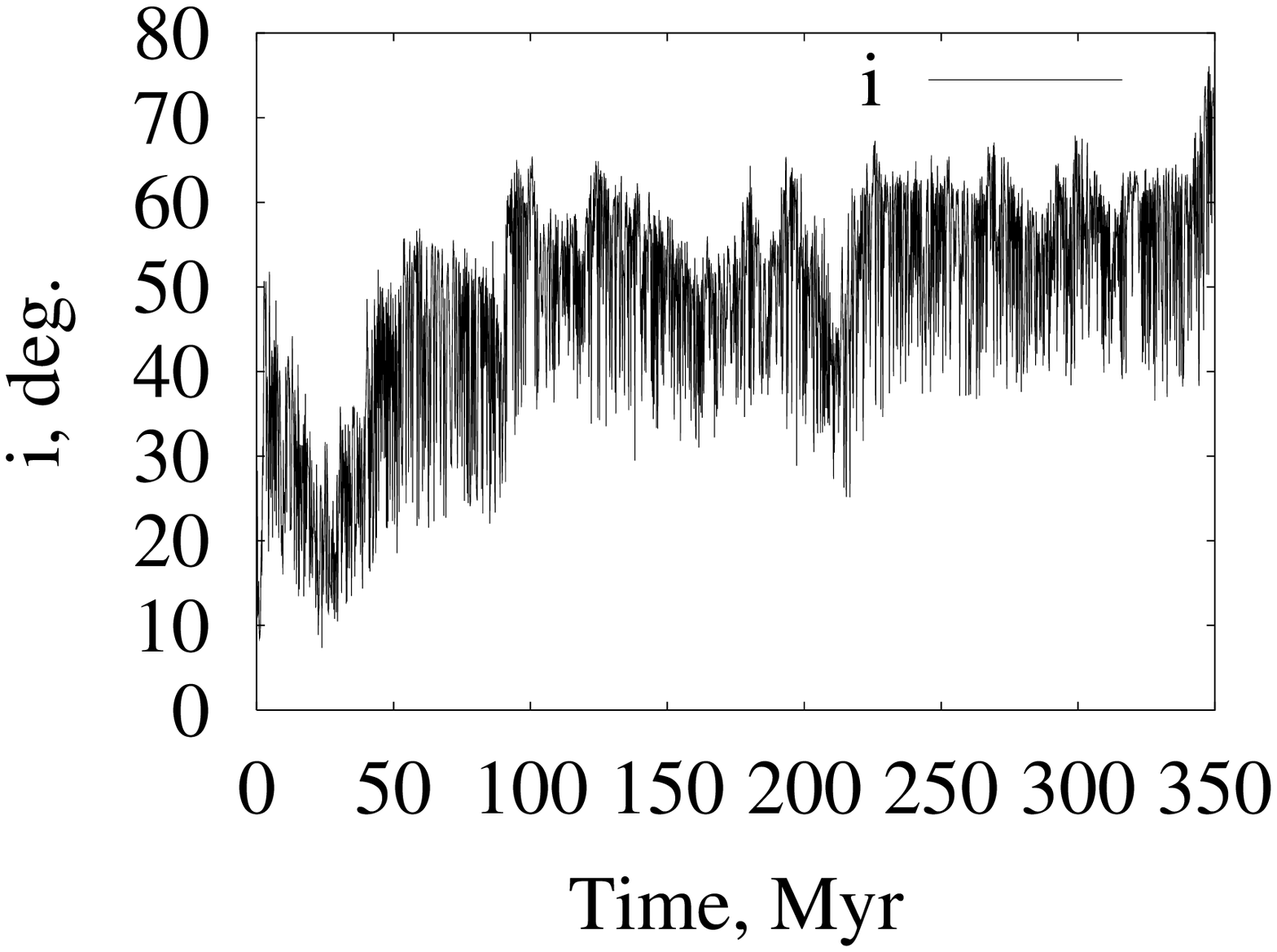}

\caption{Time variations in $a$, $e$, $q$, $Q$, 
and $i$
for a former JCO 
in initial orbit close to that of Comet 10P (a), 2P 
(b). 
For (a) at $t$$<$$0.123$ Myr $Q$$>$$a$$>$1.5 AU. 
Results from 
BULSTO code at $\varepsilon \sim 10^{-9}-10^{-8}$. 
}

\end{figure}%

     In Fig. 2 we present the time in Myr during which objects 
had semi-major axes in the interval with a width of 0.005 AU 
(Figs. 2a-b) or 0.1 AU (Figs. 2c-d). At 3.3 AU (the 
2:1 resonance with Jupiter) there is a gap for asteroids that 
migrated from the 5:2 resonance and for former JCOs (except 2P).

\begin{figure}
\includegraphics[width=91mm]{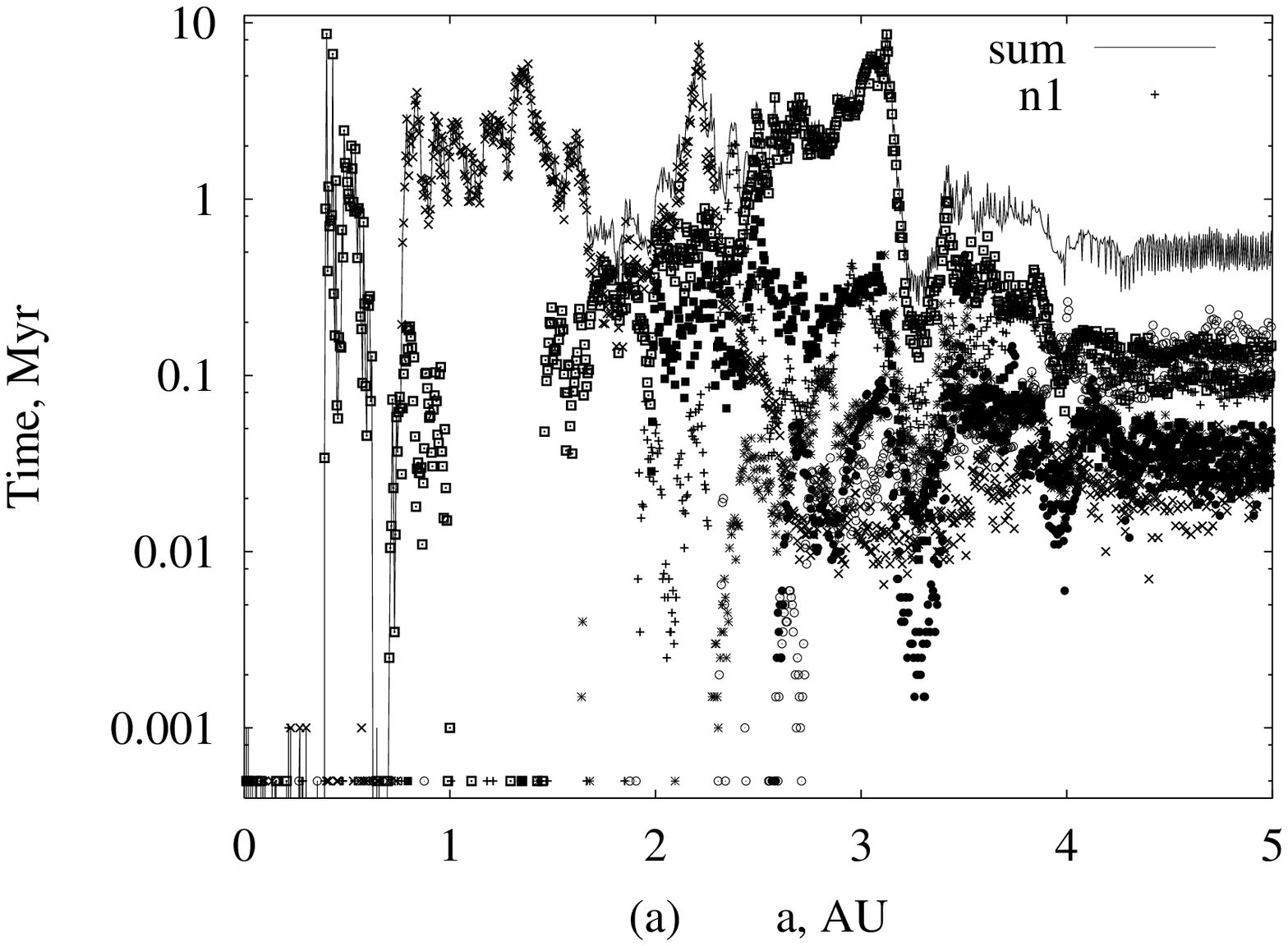}
\includegraphics[width=91mm]{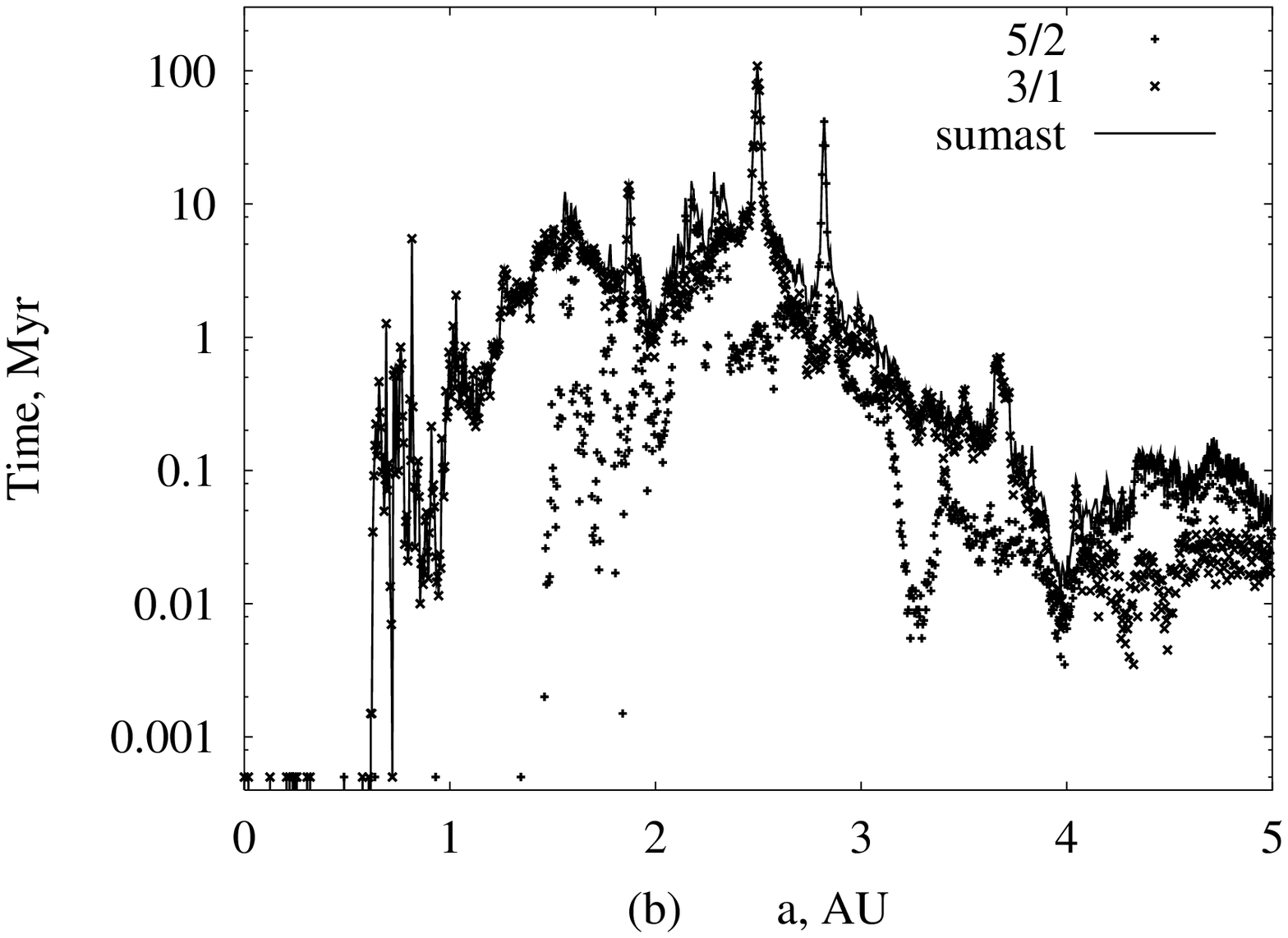} 
\includegraphics[width=91mm]{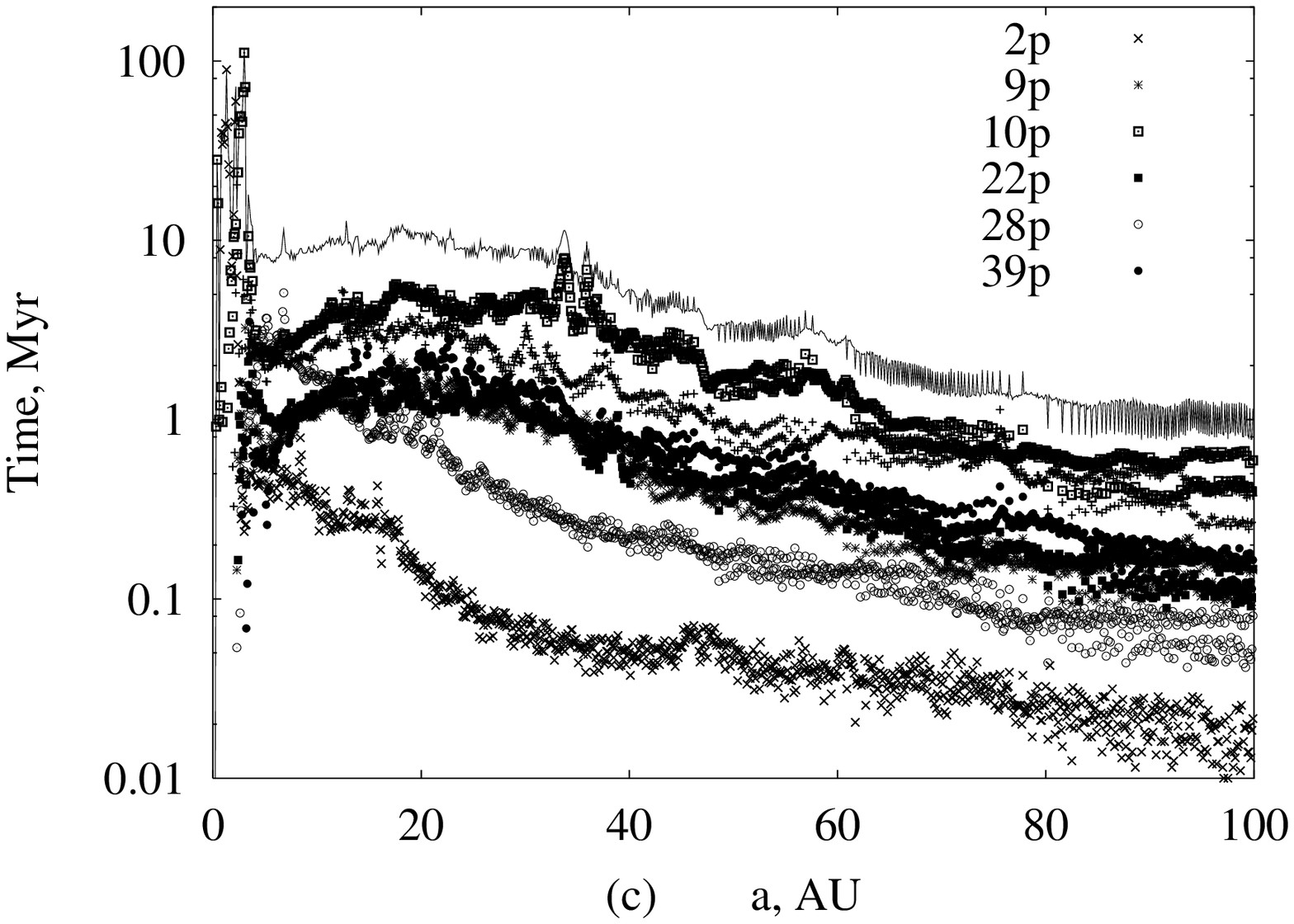}
\includegraphics[width=91mm]{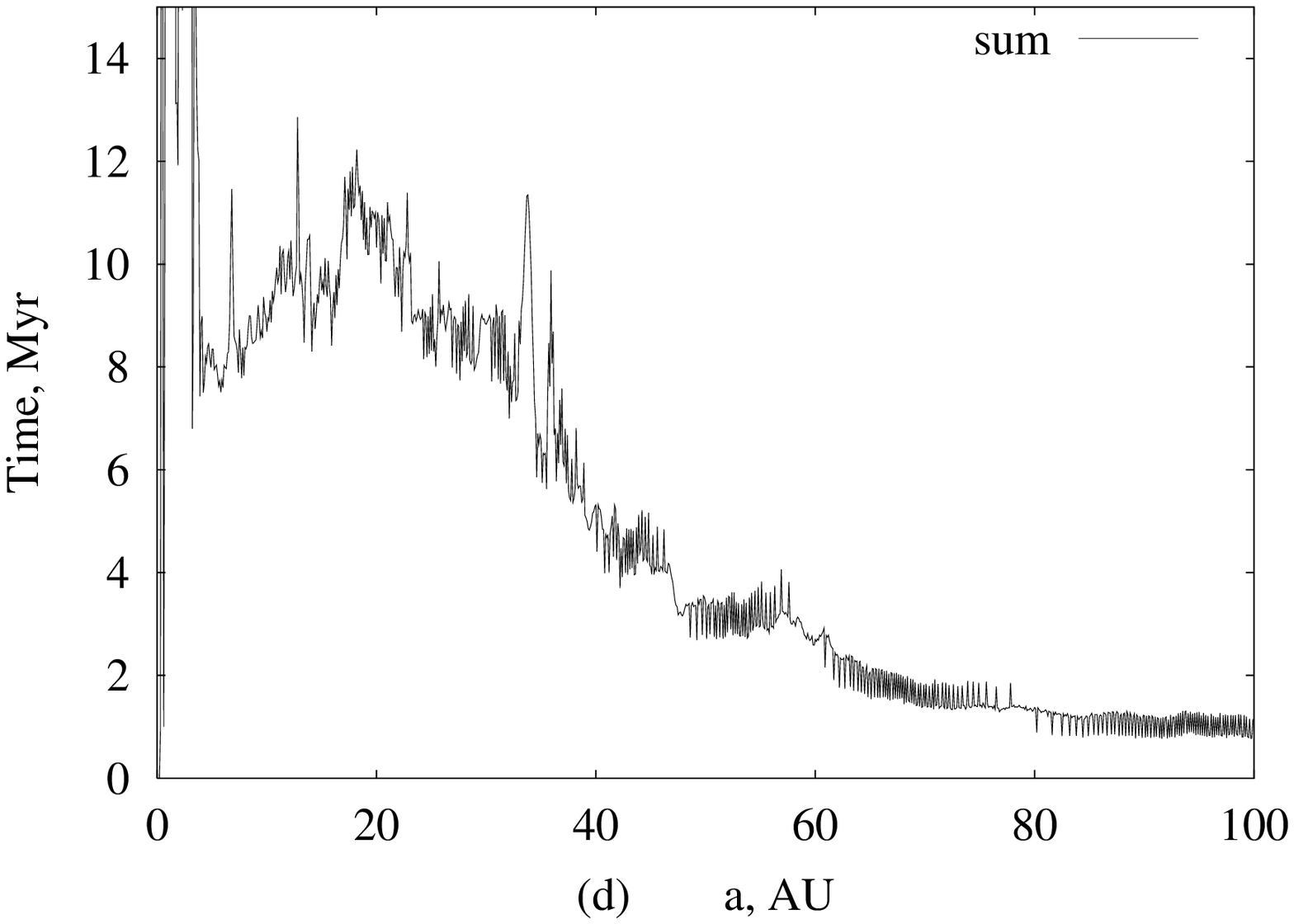}
\caption{ Distribution of migrating objects with their semi-major axes.
The curves plotted in (b) at {\it a}=40 AU  are (top-to-bottom) for sum,
10P, n1, 39P, 22P, 9P, 28P, and 2P. For Figs. (a) and (c), designations are the
same. Results from BULSTO code at $\varepsilon \sim 10^{-9} - 10^{-8}$.} 
\end{figure}%

\begin{figure}

\includegraphics[width=60mm]{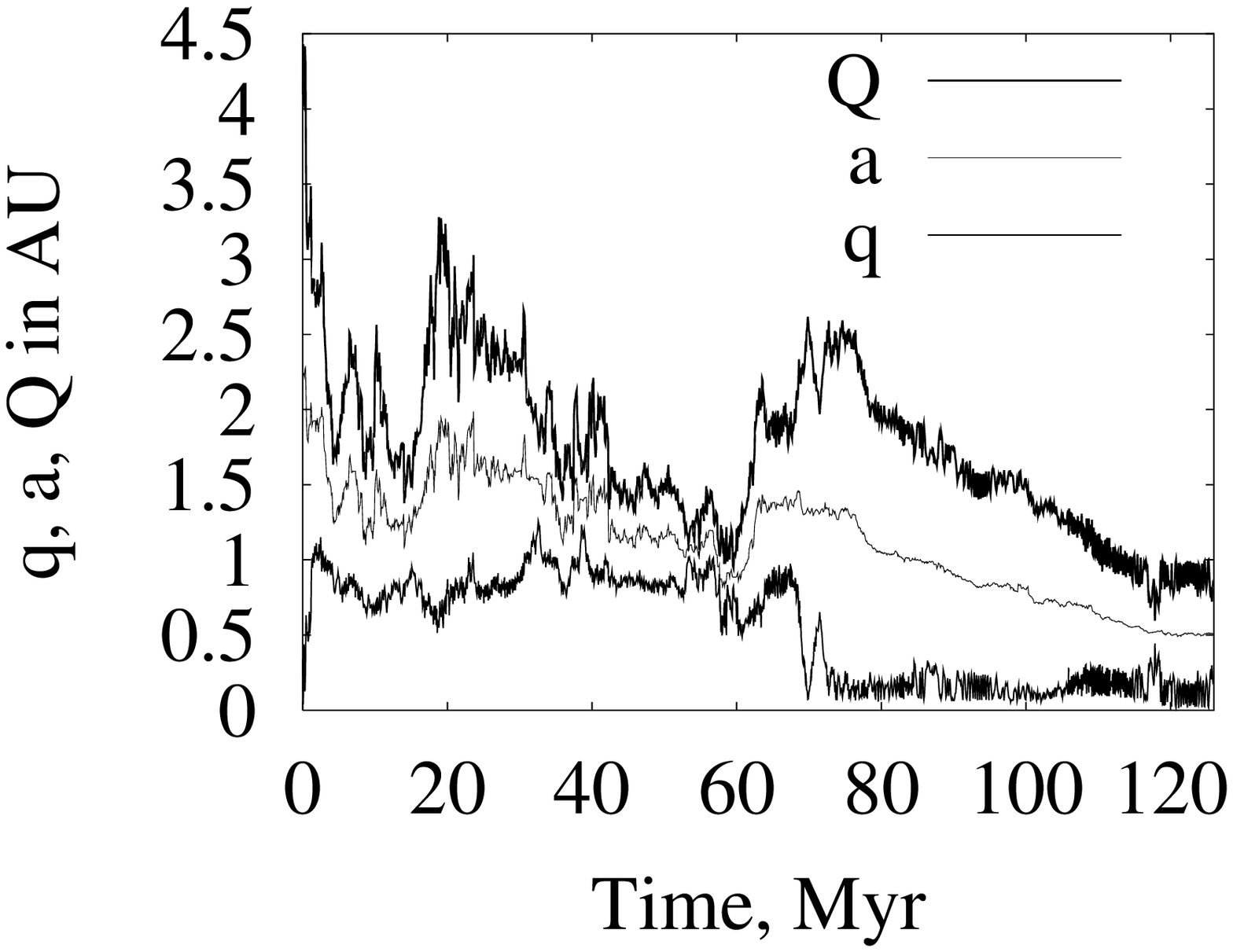}
\includegraphics[width=60mm]{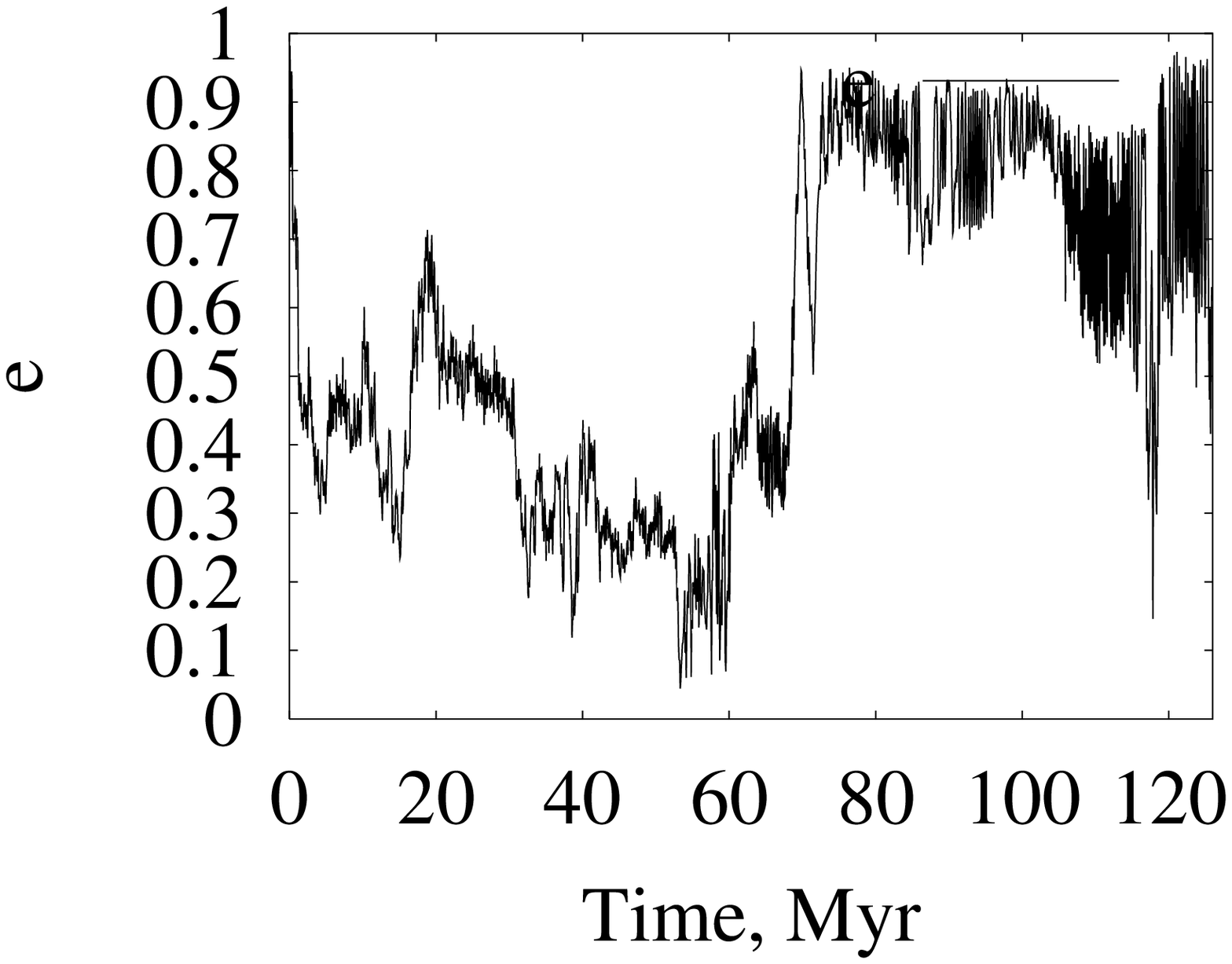}
\includegraphics[width=60mm]{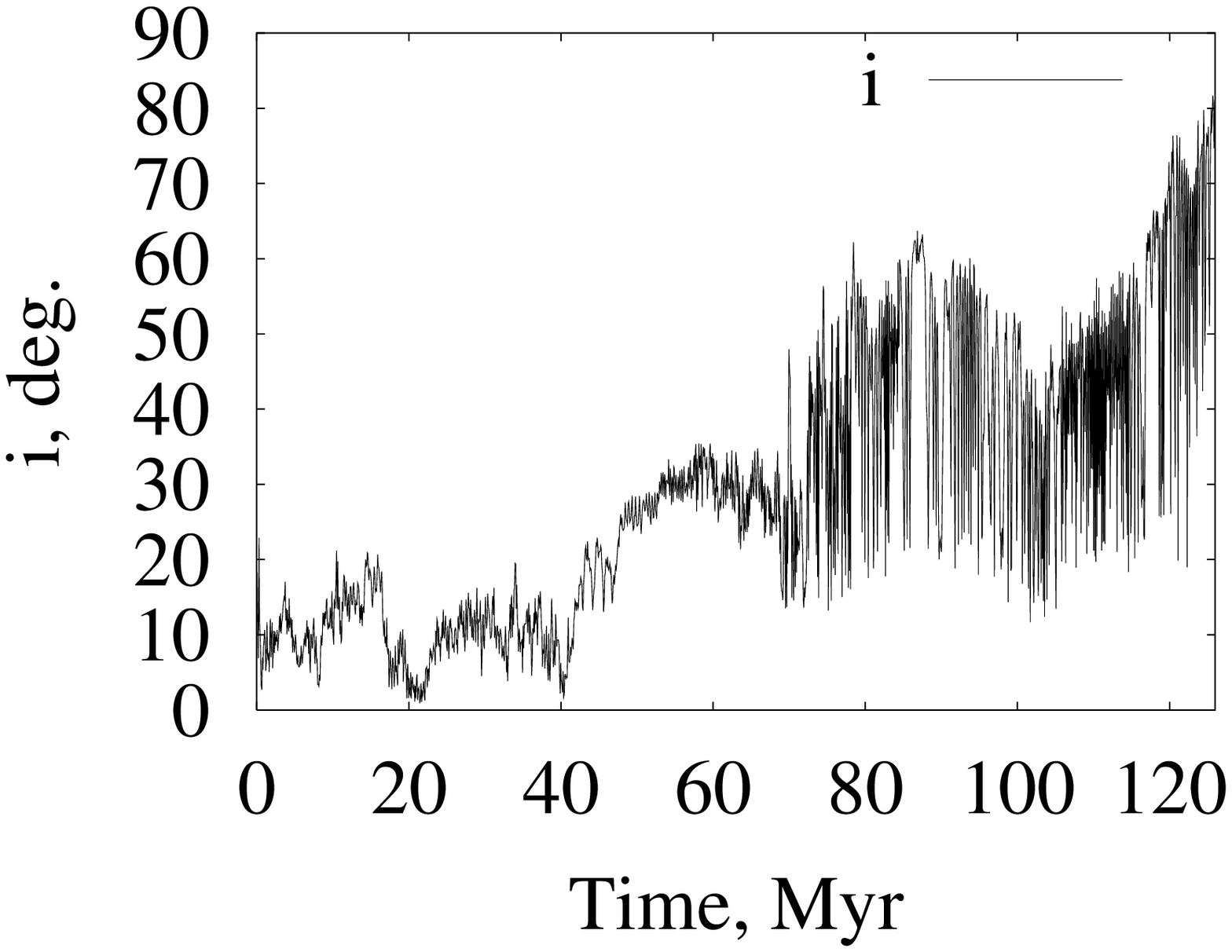}

\caption{Time variations in $a$, $e$, $q$, $Q$, 
and $i$
for a former JCO 
in initial orbit close to that of Comet 2P. 
Results from 
a symplectic method at $d_s$=10 days.}

\end{figure}%

\begin{figure}
\includegraphics[width=91mm]{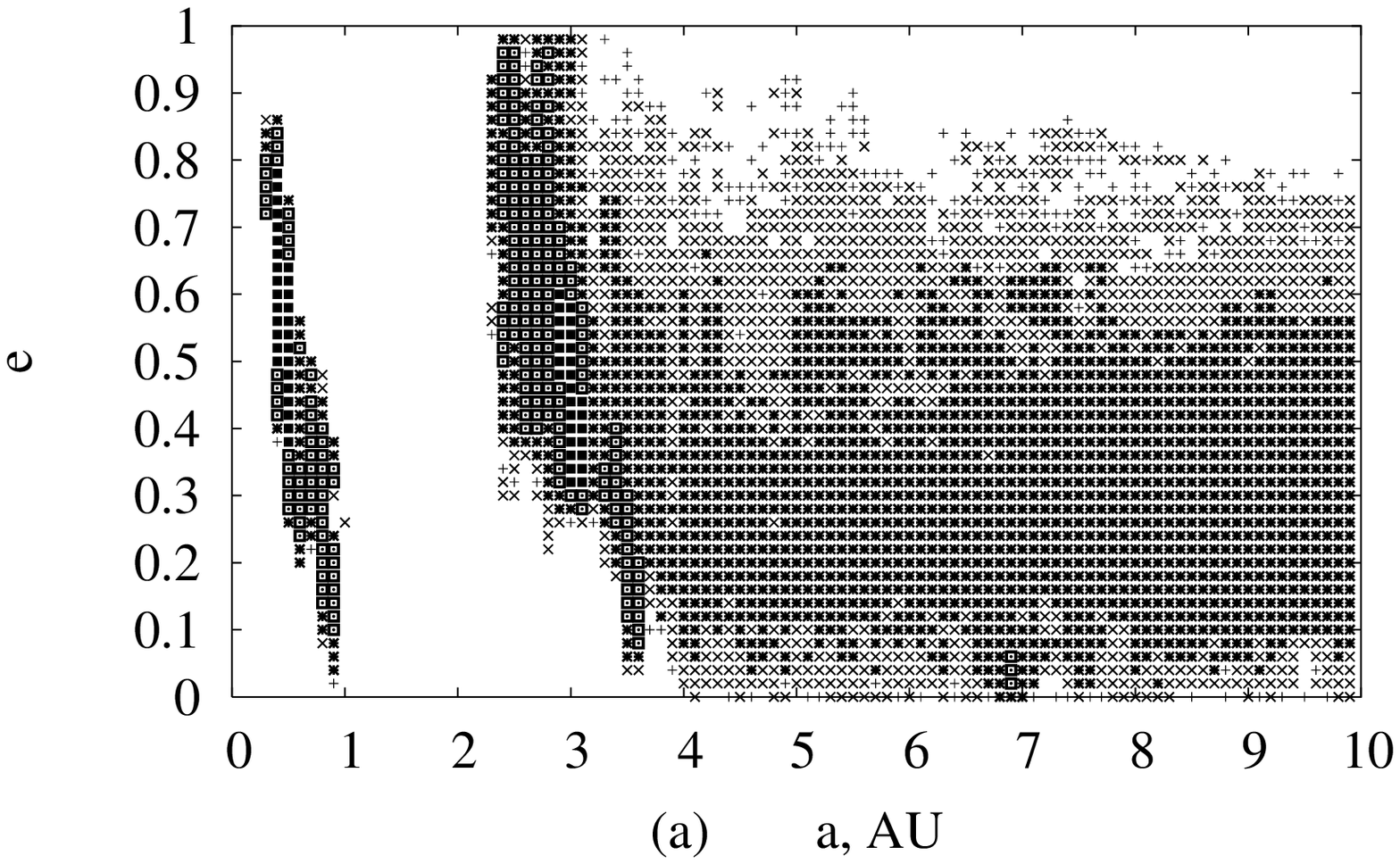}
\includegraphics[width=91mm]{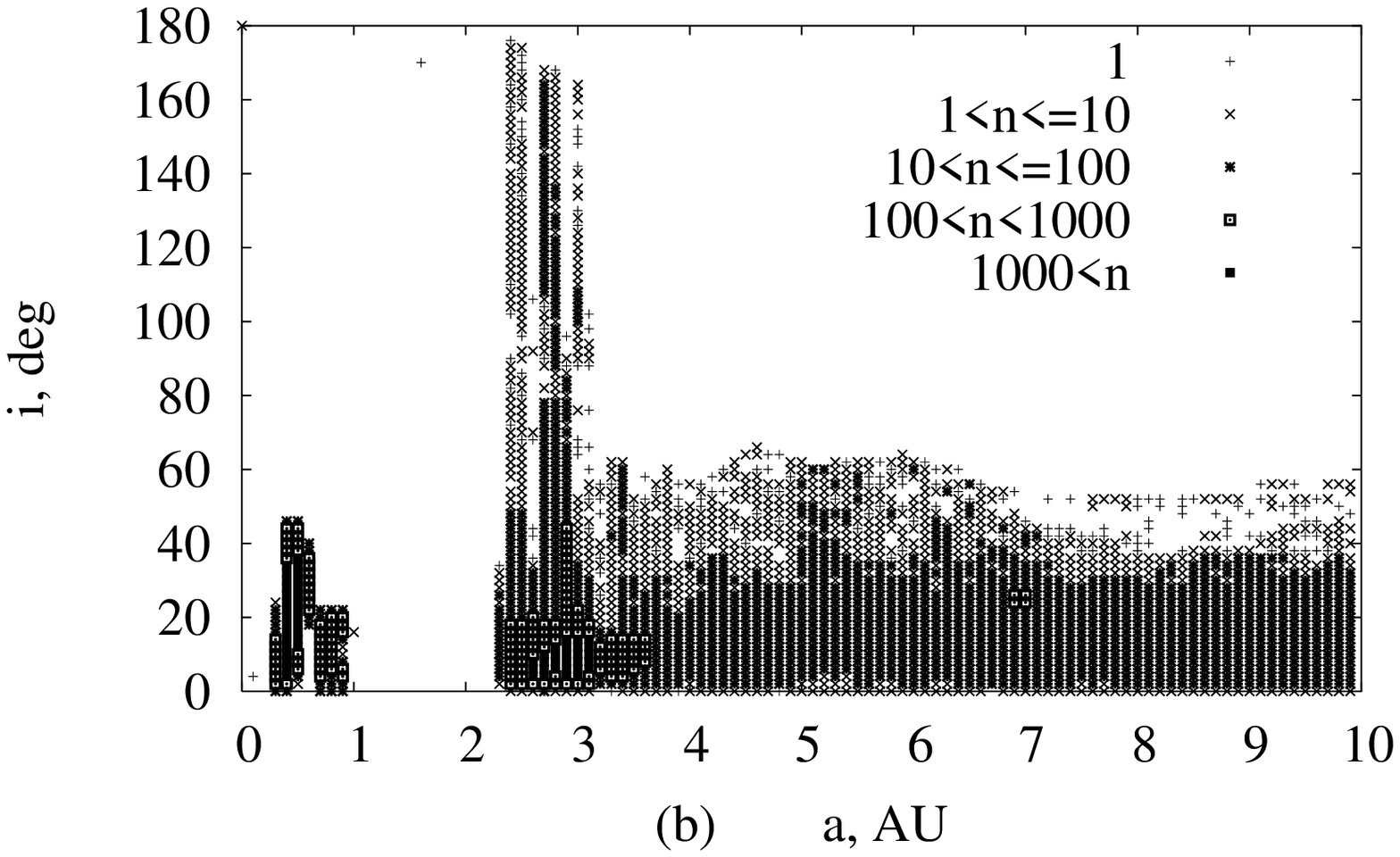}
\includegraphics[width=91mm]{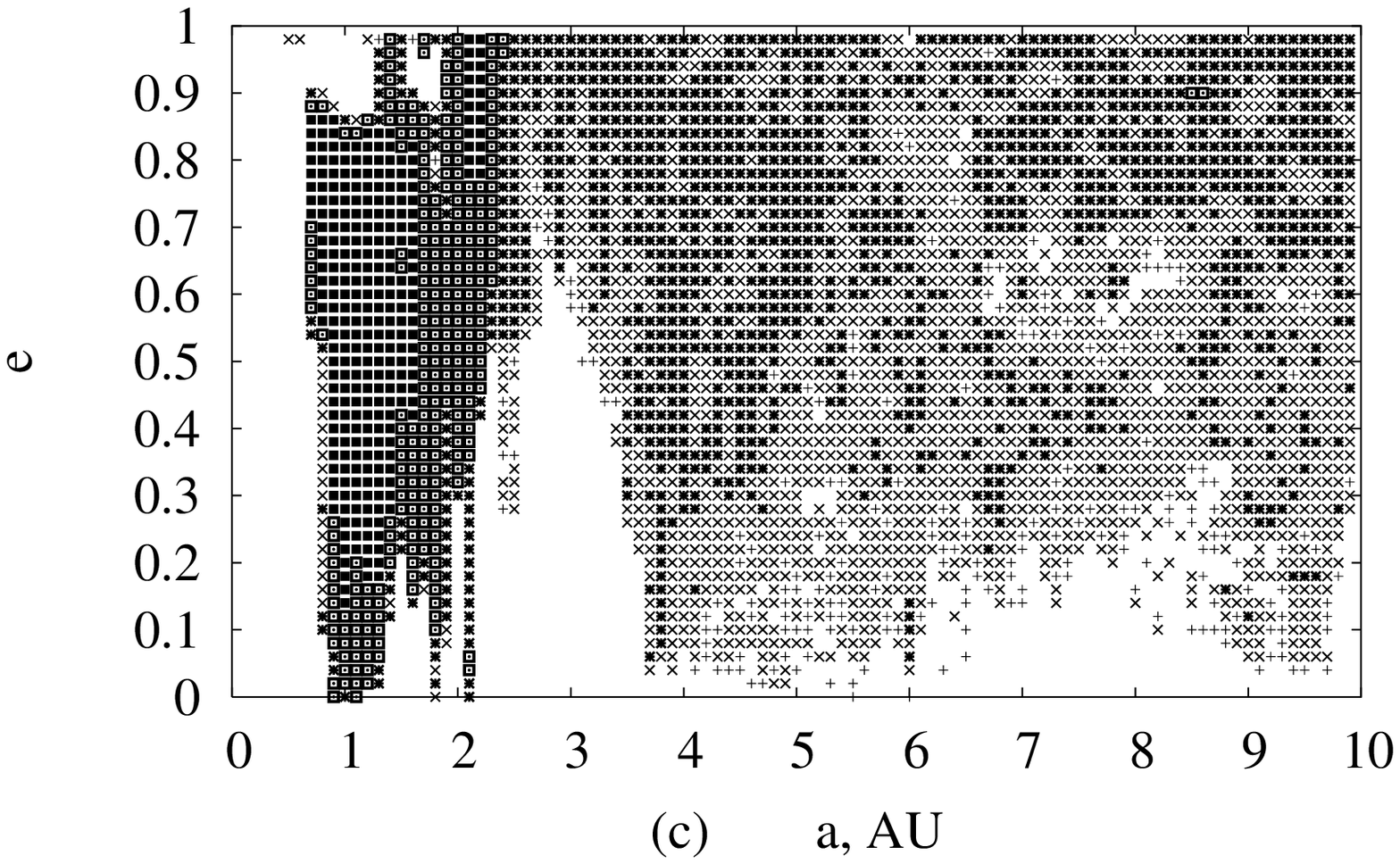}
\includegraphics[width=91mm]{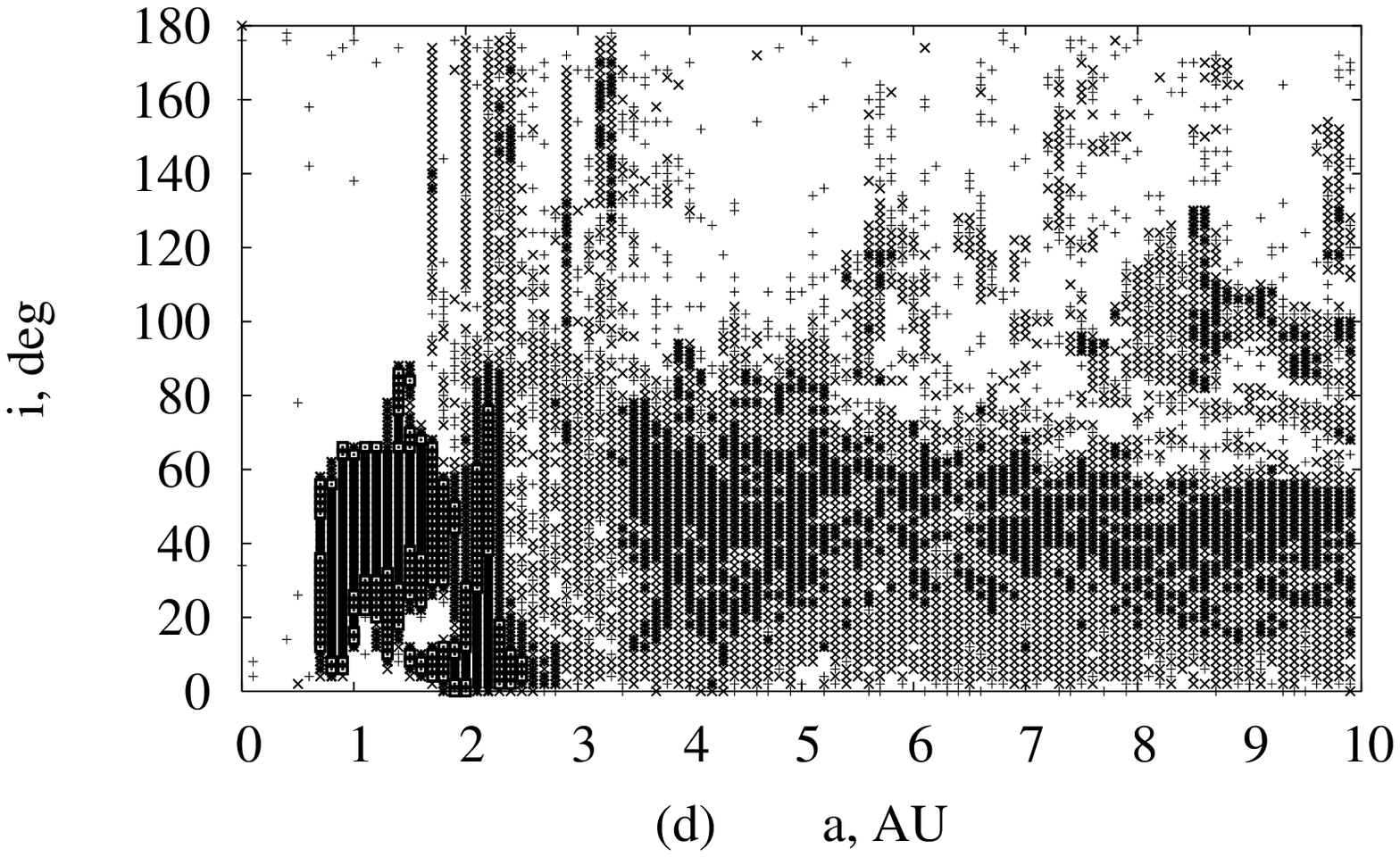}
\includegraphics[width=91mm]{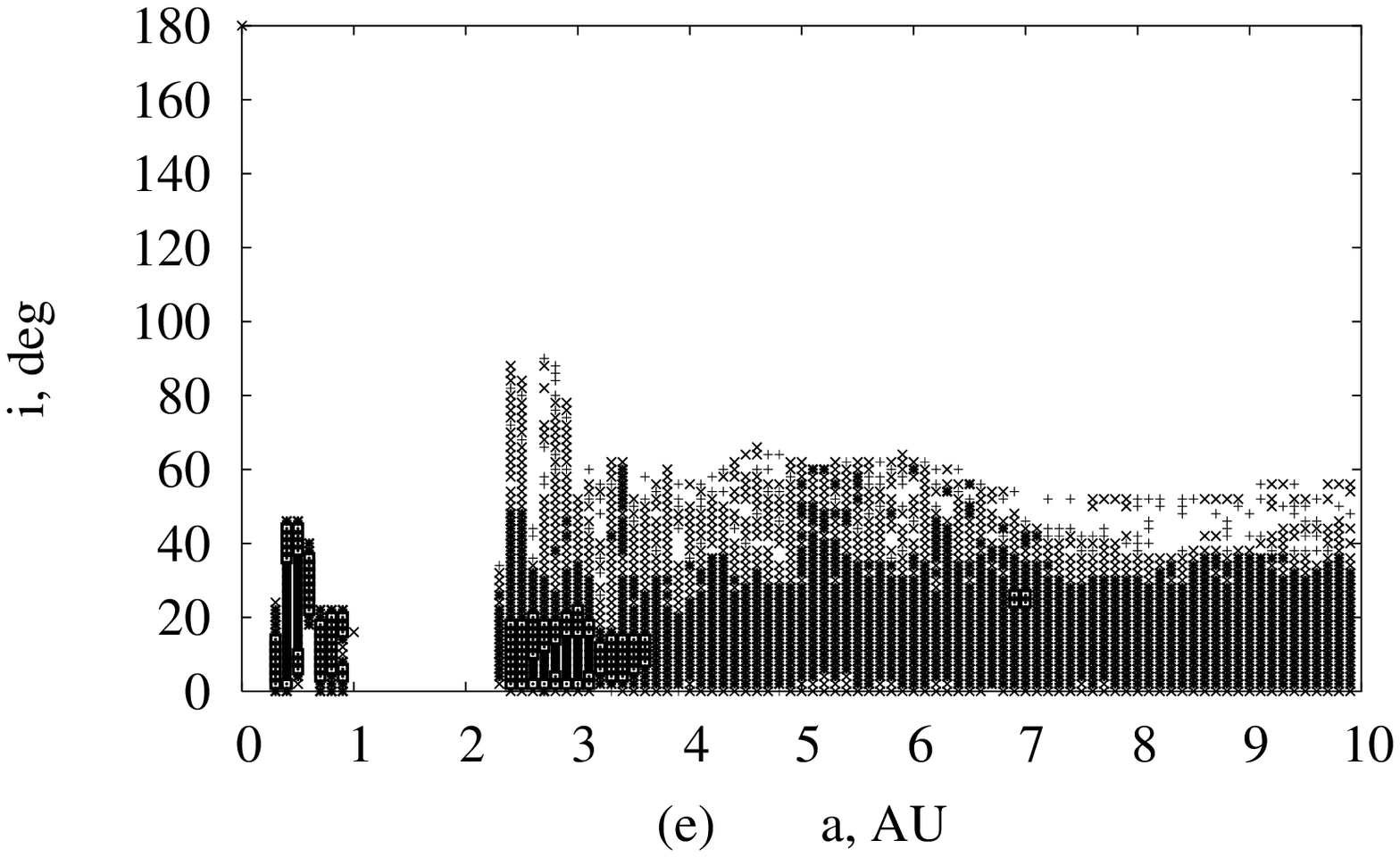}  
\includegraphics[width=91mm]{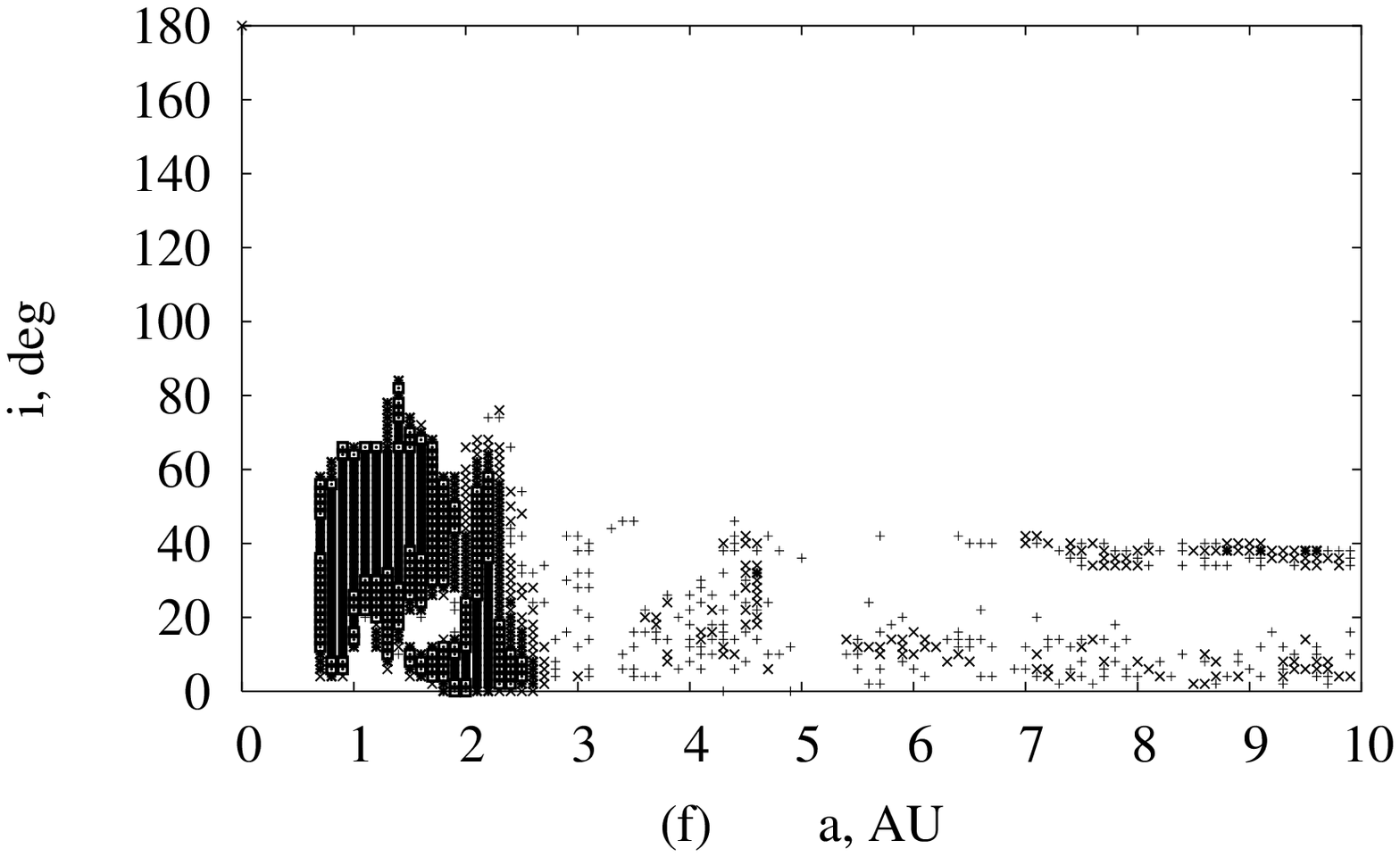}  
\includegraphics[width=91mm]{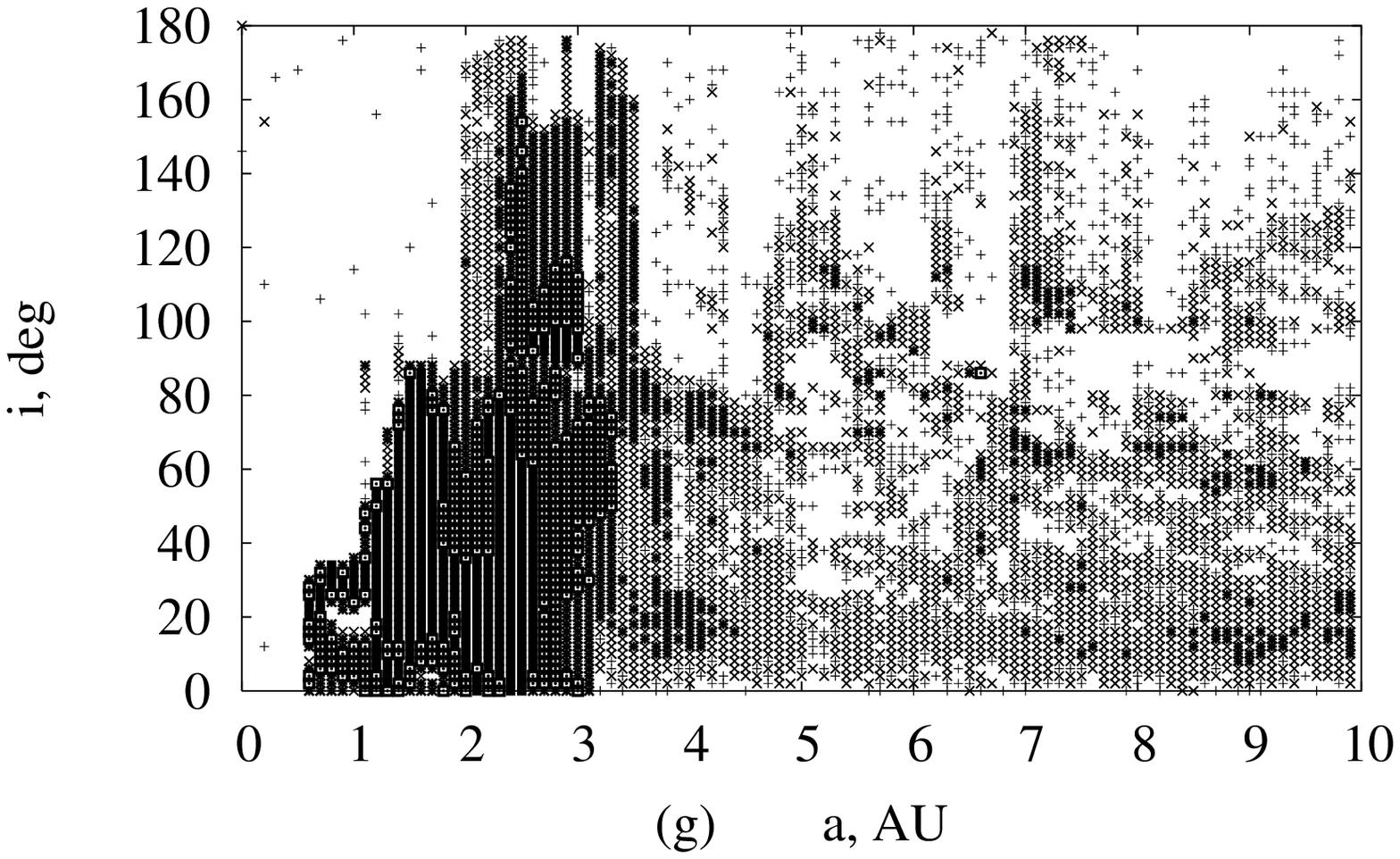}
\includegraphics[width=91mm]{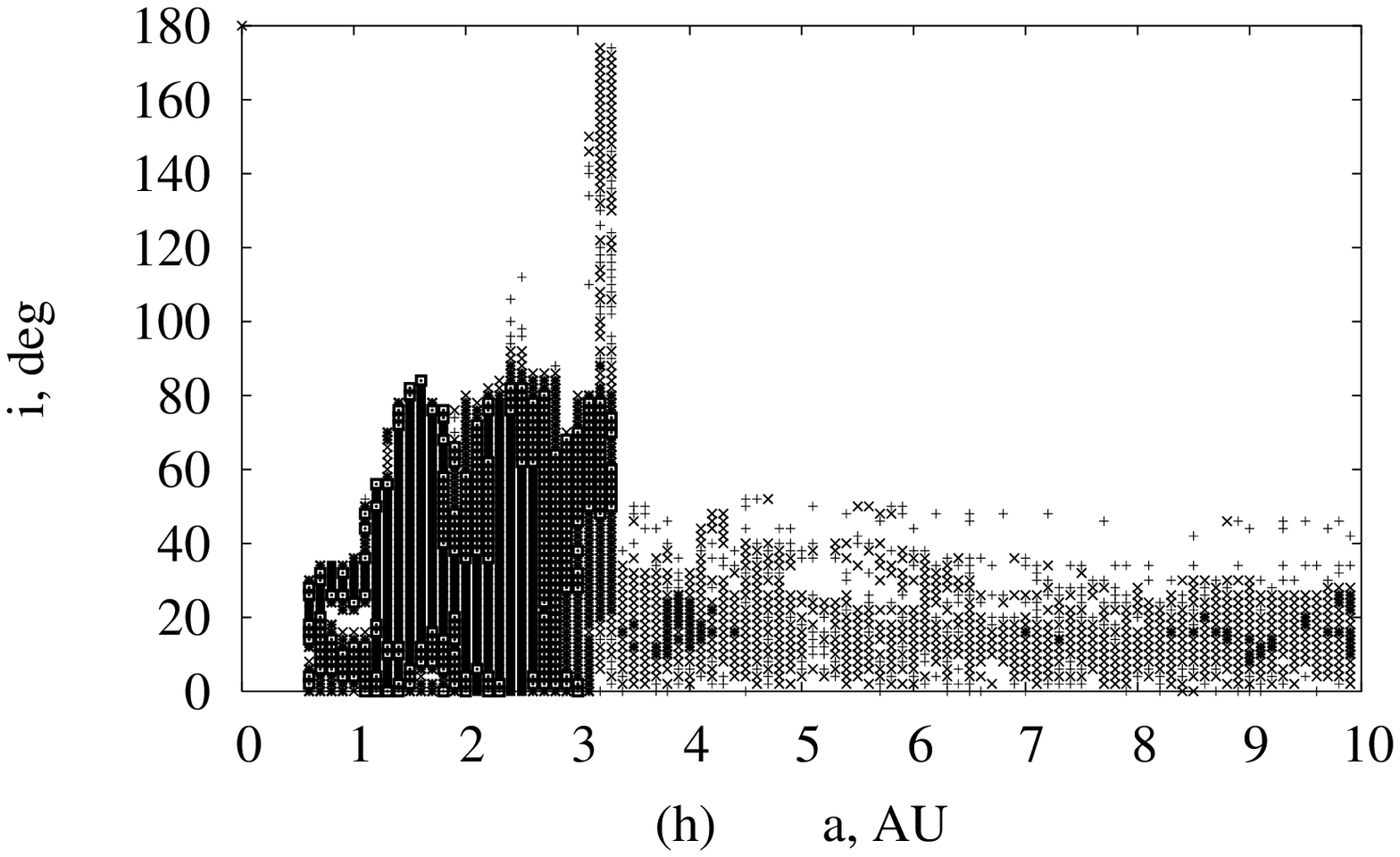}
\caption{Distribution of migrating objects in semi-major axes,
eccentricities, and inclinations for objects in initial orbits close to that of 10P (a-b,e),
2P (c-d,f), and 
the resonance 3:1 with Jupiter (g-h).
BULSTO code at $\varepsilon \sim 10^{-9} - 10^{-8}$.
For (e-f,h) it was considered that an object disappeared when perihelion distance became
less than 2 radii of the Sun. In other cases, objects disappered when they collided with the Sun. 
} 
\end{figure}%

      For the $n1$ data set, $T_J$=0.12 Myr and, while moving in 
Jupiter-crossing orbits, objects had orbital periods $P_a$$<$$10$, 
10$<$$P_a$$<$20, 20$<$$P_a$$<$50, 50$<$$P_a$$<$200 yr for 11\%, 21\%, 
21\%, and 17\% of $T_J$, respectively. Therefore, there are three 
times as many  JCOs as  Jupiter-family comets (for which $P_a$$<$20 
yr).
      We also found that some JCOs, after residing in orbits with 
aphelia deep inside Jupiter's orbit,
transfer for tens of Myr to the trans-Neptunian region, either in low 
or high eccentricity orbits. We conclude that some of the main belt 
asteroids may reach typical TNO orbits, and then become 
scattered-disk objects having high eccentricities, and vice versa. 
The fraction of objects from the 5:2 resonance that collided with the 
Earth was only 1/6 of that for the 3:1 resonance. Only a small 
fraction  of the asteroids
from the 5:2 resonance reached $a$$<$2 AU. 

The distributions of migrating former JCOs 
and resonant asteroids 
in $a$ and $e$  and in $a$ and $i$  are presented in Fig. 4. 
For each picture we considered 250 migrating objects
(288 for Fig. 4g-h), 100 intervals for $a$, and about 50--90
intervals for $e$ and $i$. Different designations correspond to different 
number $n$ of orbital elements (calculated with a step of 500 yr) which 
get in one bin (in Fig. 4 $`$$<$=$`$ means $\le$). 
All considered former JCOs very rarely reached low eccentricity orbits 
with 2$<$$a$$<$3.5 AU and 11$<$$a$$<$28 AU. 
There were many positions of objects when their perihelia were close to a 
semi-major axis of a giant planet, mainly of Jupiter.
The pictures are different for different runs.

\section*{COMPARISON OF ORBIT INTEGRATORS}

To determine the effect of the choice of orbit integrators and 
convergence criteria, we made additional runs with BULSTO 
at $\varepsilon$=10$^{-13}$ and $\varepsilon$=10$^{-12}$ 
and with a 
symplectic integrator. 
The orbital 
evolution of 5400 JCOs was computed with the RMVS3 code. 
For the  symplectic method we used an 
integration step $d_s$ of 3, 10, and 30 days.
We find that for the purposes of this paper, 
the differences between integrator choices 
(at $d_s$$\le$10 days)
are comparable to the differences between 
runs with slightly different initial conditions.  Our interpretation 
is that 1) very small numbers of particles contribute most of the 
collision probabilities
with the terrestrial planets, 2) runs with larger numbers of particles are 
more reliable, and 3) small differences in initial conditions or in 
the errors of the orbit integrators modify the trajectories 
substantially.

To illustrate these points, 
Tables 4-5  present the results obtained by
BULSTO at $\varepsilon$$\le$$10^{-12}$ and the symplectic method 
at $d_s$$\le$10 days.
Most of the results obtained with these values of $\varepsilon$ and 
$d_s$ are statistically
similar to those obtained for $10^{-9} $$\le$$ \varepsilon$$\le$$ 
10^{-8}$. For example,
a few objects spent millions of years in Earth-crossing orbits 
inside Jupiter's orbit (Figs. 1 and 3), and their probabilities of collisions with 
the Earth were thousands of times greater than for more typical 
objects.
For series $n1$ with RMVS3, the
probability of a collision with Earth for 
one object with initial orbit close to that of Comet 44P
was 88.3\%
of the total probability for 1200 objects from this series, and the 
total probability for 1198 objects was only 4\%.
This object and the object presented in Fig. 3 and in the first line of Table 4
were not included in Table 5 with $N$=1199 for $n1$ and with $N$=250 for 2P, 
respectively.
For the resonance 3:1 at $d_s$=10 days, 142 objects spent 140 and 
84.5 Myr in IEO and Aten orbits, respectively, even longer than
for $\varepsilon$$\sim$$10^{-9}$-$10^{-8}$.
Additionally, up to 40 Myr and 20 Myr were spent in such orbits by 
two other objects which had estimated probabilities of collisions 
with the terrestrial planets greater than 1 for the calculated sets 
of orbital elements. For the 2P runs 
at $\varepsilon$$\le$$10^{-12}$ and $N$=100, the 
calculated objects spent 5.4 Myr
in Apollo orbits with $a$$<$2 AU.

\begin{table}[h]

\begin{center}
\begin{minipage}{15.7cm}
\caption{Times (Myr) spent by three objects in various 
orbits, and probabilities of collisions
 with Venus ($p_v$), Earth 
($p_e$), and Mars ($p_m$)
during their lifetimes $T_{lt}$ (in Myr). Symplectic method.}
$ \begin{array}{lllll lllll l}
\hline

   & d_s $ or $ \varepsilon&$IEOs$& $Aten$&  $Al2$& $Apollo$ & $Amor$ 
& T_{lt} &p_v&p_e&p_m\\
\hline
$2P$ & 10^d & 12 & 33.6 & 73.4 & 75.6 & 4.7 & 126 &0.18&0.68&0.07\\
$44P$ & 10^d&  0 & 0    & 11.7 & 14.2 & 4.2 & 19.5 &0.02&0.04&0.002\\
$resonance $ 3:1 & 10^{-12} & 0 & 0 & 20 & 233.5 & 10.4 & 247 
&0.008&0.013&0.0007\\

\hline

\end{array}$
\end{minipage}
\end{center}
\end{table}



The values of $P_r$ presented in Table 5 are usually of the same order of
magnitude as those in Table 2, and the difference between the data 
presented in these tables is comparable to the differences
between different runs belonging to a series.
For Earth and Venus, the values of $P_r$ presented in both tables are
about 1--4 for 9P, 22P, 28P and 39P.  For 28P and 39P with the 
symplectic method
$P_r$ is about twice that for BULSTO. For 10P,  $P_r$ is 
several times larger than  for the above series, and 
for 2P it is several times larger than for 
10P. For the $n1$  run, $P_r$$>$4 for Earth.
The ratio of $P_r$ to the mass of the planet was typically several 
times larger for Mars than for Earth and Venus. The main difference 
in $P_r$ was found for the 3:1 resonance.
In this case greater values of $P_r$ were obtained for $d_s$=10 days. 
As noted above, a few exceptional objects dominated  the 
probabilities, and 
for the 3:1 resonance two objects, which had collision probabilities 
for calculated sets of orbital elements
greater than unity for the terrestrial planets, were not included 
in Table 5.  These two objects can  increase the total value of $P_r$ 
for Earth by a factor of several.


In Tables 2 and 5 we present   the mean time $T_d$ (in Kyr) 
spent in orbits with $Q$$<$4.2 AU. It can differ by three orders of magnitude
for different series of runs. In Table 5
in [ ] we present the number $N_d$ of objects each of which got $Q$$<$4.2 AU
during at least 1000 yr, the second number $N_d$ in [ ] means the same 
but only for $q$$<$1.017 AU. For series 2P and the 3:1 resonance with Jupiter
both values of $N_d$
were almost the same, for 10P they differed by a factor of
2.9 (symplectic runs) or 3.7 (BULSTO), and for other series only a small
portion of decoupling objects crossed the orbit of the Earth.
For most runs (except for 2P and asteroids) the number of objects
which got $Q$$<$4.7 AU was several times larger than 
that for $Q$$<$4.2 AU. 

We also did some symplectic runs with $d_s$=30 day. For most of the 
objects we got similar results, but
about 0.1\%  of the
objects    reached Earth-crossing orbits with $a$$<$2 AU for several 
tens of Myr
and even IEO orbits. These few bodies increased the 
mean value of $P$ by a factor of more than 10. With $d_s$=30 days, 
four  objects from the runs $n1$, 9P, 10P had a probability of 
collisions
with the terrestrial planets 
for calculated sets of orbital elements
greater than 1 for each, and for 2P 
there were 21 such objects among 251 considered.
For resonant asteroids, we also obtained much larger values than 
those with BULSTO for  $P$ and $T$ with RMVS3 at
$d_s$=30 days, and similarly  for the 3:1 resonance even at $d_s$=10 
days. For this resonance it may be
better to use $d_s$$<$10 days.
Probably, the results of symplectic runs at $d_s$=30 days can be considered as 
such migration that includes some nongravitational forces.


\begin{table}[h]
\begin{center}
\begin{minipage}{18cm}

\caption{
Probabilities of collisions with the terrestrial planets.
Designations are same as those for Table 2.
$T_d$ is the mean time (in Kyr) spent in orbits with $Q$$<$4.2 AU;
in [ ] we present the number $N_d$ of objects each of which got $Q$$<$4.2 AU
during at least 1000 yr, the second number in [ ] means the same 
but only for $q$$<$1.017 AU.
Results from
the BULSTO code at $\varepsilon$$\le$$10^{-12}$ and  a symplectic method
at $d_s$$\le$10 days.}

$ \begin{array}{lllcc ccccc ccl l}

\hline

    &&&$V$ & $V$ & $V$ & $E$ & $E$ & $E$ & $M$ & $M$ & $M$ &  &\\


  &\varepsilon $ or $ d_s& N&P_r & T &T_c& P_r & T &T_c& P_r & T &T_c& 
r,$ $$ $$ $$ $ T_J & T_d [N_d]\\

\hline

n1&\le$$10^d&1200&25.4 & 13.8 &0.54& 40.1 & 24.0 &0.60&2.48 & 35.2 
&14.2& 3.0 $  $117 & 25.7 [41/3]\\
n1&\le$$10^d&1199&7.88 & 9.70 &1.23& 4.76 & 12.6 &2.65&0.76 & 16.8 
&22.1& 2.8 $  $ 117 &10.3 [40/2]\\
$2P$&\le$$10^{-12}&100 &321&541&1.69&146&609&4.2   &14.8 & 634 &42.8 & 27. $  $ $ $20 & 247 [100/100]\\
$2P$&10^d & 251 &860&570& 0.66 & 2800 &788 & 0.28 &294 & 825 & 2.81& 22. $  $$ $0.29&614 [251/251]\\
$2P$&10^d & 250 &160&297& 1.86 & 94.2 & 313 & 3.32 & 10.0&324 &32.3&35. $  $$ $0.29& 585 [250/250]\\
$9P$&10^d & 400 &1.37&3.46&2.53&3.26&7.84&2.40&1.62&23.8&14.7& 1.1 $  $$ $128& 8.0 [13/3]\\
$10P$&\le$$10^d &450&14.9&30.4&2.04&22.4&41.3&1.84&6.42& 113.&17.6&1.5 $  $$ $85&44. [70/24]\\
$22P$&10^d & 250& 0.68 &2.87&4.23&1.39&4.96&3.57&0.60 & 11.5&19.2& 1.5 $  $$ $121 & 0.6 [3/0]\\
$28P$&10^d & 250 & 3.87 & 35.3& 9.12& 3.99 & 59.0 &14.8& 0.71 & 109. 
&154&2.2 $  $$ $535 &3.3 [7/0]\\
$39P$&10^d & 250 & 2.30 &2.68& 1.17 &2.50& 4.22 & 1.69 & 0.45 & 7.34& 
16.3 & 2.2 $  $$ $92 & 0.5 [2/0]\\
3:1&\le$$10^{-12}& 70 & 1162 &1943&1.67& 1511& 5901 &3.91& 587& 
803&1.37& 4.6 $  $$ $326 & 8400 [70/70]\\
3:1&10^d& 142 &27700&8617&0.31& 2725& 9177& 
3.37& 1136&9939&8.75& 16. $  $$ $1244&5000 [142/140]\\
5:2&10^{-12}& 50 & 130   &113&0.87& 168 & 230 &1.37&46.2 &507 
&11.0& 1.4 $  $$ $166 &512 [50/4]\\
5:2&10^d& 144 & 58.6&86.8&1.48&86.7&174 &2.01&17.&355&20.9&1.7 $  $$ $224& 828 [144/13]\\

\hline

\end{array} $
\end{minipage}
\end{center}
\end{table}

      In the case of close encounters with the Sun (Comet 2P and 
resonant asteroids), the probability
$P_S$ of collisions with the Sun  was larger for RMVS3 than for BULSTO, 
and for
$10^{-13} $$\le$$ \varepsilon $$\le$$ 10^{-12}$ it was greater than 
for $10^{-9}$$ \le$$ \varepsilon $$\le$$ 10^{-8}$
($P_S$=0.75 for the 3:1 resonance 
at $d_s$=3 days). This probability is presented below for several runs:

\begin{table}[h]

\begin{center}
\caption{Probability of collisions with the Sun}
$ \begin{array}{lllllll}
\hline
  & \varepsilon=10^{-13} & \varepsilon=10^{-12} & \varepsilon=10^{-9} 
& \varepsilon=10^{-8} & d_s=10 $ days$ & d_s=30 $ days$ \\
\hline
$Comet 2P$& 0.88 & 0.88 & 0.38 & 0.32& 0.99 & 0.8 \\

$resonance $ 3:1 &0.46& 0.5& 0.156 & 0.112 & 0.741 & 0.50 \\

$resonance $ 5:2 & &0.06& 0.062 & 0.028 & 0.099 & 0.155\\
\hline
\end{array} $

\end{center}
\end{table}

For Comet 2P the values of $T_J$ were much smaller for RMVS3 than 
those for BULSTO and they were smaller
for smaller $\varepsilon$; for other runs these values do not depend 
much on the method.
In the direct integrations reported by Valsecchi et al. (1995), 13 of 
the 21 objects fell into the Sun,
so their value of $P_S$=0.62 is in accordance with our results 
obtained by BULSTO; it is less than
that for $\varepsilon$=$10^{-12}$, but greater than  for 
$\varepsilon$=$10^{-9}$.
Note that even for different $P_S$ the data presented in Tables 2 and 
5 usually are similar. As we did not calculate
collision probabilities of objects with planets by direct 
integrations, but instead calculated them with the random phase 
approximation from the
orbital elements, we need not make integrations with extremely high 
accuracy. Ipatov (1988b) showed that for BULSTO
the integrals of motion were conserved better and the plots of 
orbital elements for closely separated values of  $\varepsilon$ were
closer to one another at  $10^{-9} $$\le$$ \varepsilon $$\le$$ 
10^{-8}$. The smaller the value of $\varepsilon$, the more
integrations steps are required, so $\varepsilon$$\le$$10^{-12}$ for 
large time intervals are not necessarily better than those for 
$10^{-9} $$\le$$ \varepsilon $$\le$$ 10^{-8}$. Small $\varepsilon$ is 
clearly
necessary for close encounters. Ipatov and Hahn (1999) and Ipatov 
(2000) found that former JCOs
reached resonances more often for BULSTO than for RMVS3 at $d_s$=30 days. Therefore we
made most of our  BULSTO runs with  $10^{-9} $$\le$$ \varepsilon 
$$\le$$ 10^{-8}$.
For a symplectic method it is better to use smaller $d_s$ at a 
smaller distance $R$ from the Sun,
but in some runs  $R$ can vary considerably during the evolution.

W. Bottke pointed out that H. Levison showed that
it is difficult to detect solar collisions in any numerical integrator,
so he removed objects with $q$$<$$q_{\min}$. The results presented above
were obtained considering collisions with the Sun, but we also
investigated what happens if we consider $q_{\min}$ equal to $k_S$
radii $r_S$ of the Sun. For $k_S$=2, some results are
presented in Fig. 4e-f,h. The only difference with the runs
that considered collisions with the Sun
is that  for those runs
for series 2P and 10P and for the 3:1 resonance, 
some objects reached $90^\circ$$<$$i$$<$$180^\circ$ (mainly with 
2$<$$a$$<$3.5 AU) (Fig. 4b,d,g). For $k_S$=2 there were no comets with  
$i$$>$90$^\circ$
and there were only a few orbits of asteroids with $i$$>$90$^\circ$ 
(Fig. 4e-f,h). The consideration of $q_{\min}$ at $k_S$=3 did not
influence the collision probabilities with the terrestrial planets or
getting orbits with $a$$<$2 AU. For example, with BULSTO for the two objects with the 
largest collision probabilities, the time spent in orbits with
$a$$<$2 AU decreased by only 0.3\% for 2P at $k_S$=3 and was the same
for 10P at $k_S$=10. 


\section*{MIGRATION FROM BEYOND JUPITER TO THE TERRESTRIAL PLANETS}

According to Duncan et al. (1995), the fraction $P_{TNJ}$ of TNOs reaching
Jupiter's orbit under the influence of the giant planets in 1 Gyr is
0.8-1.7\%. As the mutual gravitational influence of TNOs can play a 
lar\-ger role in variations of their orbital elements than collisions 
(Ipatov, 2001), we 
considered the upper value of
$P_{TNJ}$.
Using the total of $5\times10^9$ 1-km 
TNOs with $30$$<$$a$$<$$50$ AU,  and assuming
that the mean time for a body to move in a Jupiter-crossing orbit is 
0.12 Myr, we find that about $N_{Jo}$=$10^4$ 1-km former TNOs are now 
Jupiter-crossers, and 3000 are JFCs.
Using the total times spent by $N$
simulated JCOs in various orbits, we obtain the following 
numbers of 1-km former TNOs now moving in several types of orbits:
\begin{center}
$ \begin{array}{cccccccc}
\hline
$$N$$ &$method$& $series$ & $IEOs$ & $Aten$ & $Al2$ & $Apollo$ & $Amor$ \\
\hline
$3100$ & $BULSTO+RMVS3$ & $n1$  & 0   & 0   & 480 & 1250 & 900 \\
$1900$ & $BULSTO$ & $n1$  & 0   & 0   & 25 & 720 & 1000 \\
$7350$ &$BULSTO$ & $all without 2P$& 120 &  40 & 250 & 2500 & 1750\\
$7852$ &$BULSTO$ & $all$& 110 & 950 & 4500 & 7200 & 1880\\
\hline

\end{array} $
\end{center}

For example, the number of IEOs $N_{IEOs}$=$N_{Jo}t_{IEO}$/$(N_J t_J)$,
where $t_{IEO}$ is the total time during which $N_J$ former JCOs 
moved in IEOs' orbits, and $N_J t_J$ is the total time during which 
$N_J$ JCOs moved in Jupiter-crossing orbits.
As we considered mainly the 
runs with relatively high migration to the Earth, 
the actual 
number of NEOs
is smaller by a factor of several, the actual portion
of IEOs and Atens 
can be smaller and that for Amors
can be larger than those in the lines in the Table at $N$$>$7000. 
Even if the number of Apollo objects is 
smaller than that based on $n1$ runs,
it may still be comparable to the real number (750) of 1-km
Earth-crossing objects (half of them are in orbits with $a$$<$2 AU), 
although the
latter number does not include those in highly eccentric orbits.

The values of the characteristic time (usually $T_c$) for the 
collision of a former JCO or a resonant
asteroid with a planet (see Tables 2 and 5) are greater than the values 
of $T_f$ for NEOs in Table 1,
$T_c$$\approx$1.1 Gyr for 7852 objects,
so we expect that the mean inclinations and eccentricities
of unobserved NEOs are greater than those for the NEOs that are 
already known.   Jedicke et al. (2003) found similar results.
On average, the values of $T_c$ for our $n1$ series  and for most of 
our simulated JCOs were not greater than those for our calculated 
asteroids, and migrating Earth-crossing objects had similar $e$ and 
$i$ for  both former JCOs and resonant asteroids. 
Former JCOs, which move in Earth-crossing orbits for more than 1 Myr, 
while moving in such orbits, usually had larger $P$ and smaller $e$ and $i$
(sometimes similar to those of the observed NEOs, see Figs. 1 and 3) than other JCOs.
It is easier to observe orbits with smaller values of $e$ and $i$, and 
probably, many of the NEOs moving in orbits with large values of $e$
and $i$ have not yet been discovered.
About 1\% of the observed Apollos cross Jupiter's orbit, and an 
additional 1\% of Apollos have aphelia between 4.7-4.8 AU,
but these Jupiter-crossers are far from the Earth  most of time, so 
their actual fraction of ECOs is greater
than  for observed ECOs. The fraction of Earth-crossers among 
observed Jupiter-family comets is about 10\%.
This is a little more than $T/T_J$ for our $n1$ runs, but less than 
for 7850 JCOs.
For our former resonant asteroids, $T_J$ is relatively large 
($\approx$0.2 Myr), and such asteroids can reach cometary orbits.

Comets are estimated to be active for $T_{act}$$\sim$$10^3$--$10^4$ yr. 
$T_{act}$ is smaller for closer encounters with the Sun (Weissman et al., 2002),
so for Comet 2P it is smaller than for other JFCs.
Some former comets can move for tens or
even hundreds of Myr in NEO orbits,  so the number of extinct comets 
can exceed the number of active comets by several orders of magnitude.
The mean time spent by Encke-type objects in Earth-crossing orbits is 
$\ge$0.4 Myr (even for $q_{\min}$). 
This time corresponds to $\ge$40-400  
extinct comets of 
this type. Note that the diameter of Comet 2P 
is about 5-10 km, so the number of smaller extinct comets can be much larger.

      The above estimates of the number of NEOs are approximate. For 
example, it is possible that the number of 1-km TNOs is several times 
smaller than $5\times10^9$,
while some scientists estimated that this number can be up 
to $10^{11}$ (Jewitt, 1999). Also,
the fraction of TNOs that have migrated towards the Earth might be 
smaller. On the other hand, the above
number of TNOs was estimated for $a$$<$50 AU, and TNOs from more 
distant regions can also migrate inward.
Probably, the Oort cloud  could also supply Jupiter-family comets. 
According to Asher et al. (2001), the rate of a cometary
object decoupling from the Jupiter vicinity and transferring to an 
NEO-like orbit is increased by a
factor of 4 or 5 due to nongravitational effects
(see also Fernandez and Gallardo, 2002). This would result 
in  larger values of $P_r$ and $T$
than those shown in Tables 2 and 5.
      Our estimates show that, in principle, the trans-Neptunian belt 
can provide a significant portion of the Earth-crossing objects, 
although many NEOs clearly came from the main asteroid belt. It may 
be possible to explore former TNOs near the Earth's orbit without 
sending spacecraft to the trans-Neptunian region.
According to our results, many
former Jupiter-family comets can have orbits typical of asteroids, 
and collide with the Earth from typical NEO orbits.

      Based on the 
collision probability $P=4\times10^{-6}$ 
we find that  1-km former TNOs now collide with the Earth once in 
3 Myr. This value of $P$ is smaller
than that for our $n1$ runs, does not include the 
'champions' in collision probability, 
and is only 1/20 of that for our 7852 JCOs.
Using $P=4\times10^{-6}$ and 
assuming the total mass of planetesimals that ever crossed Jupiter's 
orbit is $\sim$$ 100m_\oplus$, where
$m_\oplus$ is the mass of the Earth (Ipatov, 1993, 2000), we conclude 
that the total mass of bodies that impacted the
Earth is $4\times10^{-4} m_\oplus$. If ices comprised only half of this mass,
then the total mass of ices $M_{ice}$ that were delivered to the 
Earth from the feeding zone of
the giant planets is about the mass of the terrestrial oceans 
($\sim2\times10^{-4} m_\oplus$).

      The calculated probabilities of collisions of objects with 
planets show that the fraction of the mass of the planet  delivered 
by short-period comets can be greater for Mars and Venus than for the 
Earth 
(compare the values of $P$$/$$m_{pl}$
using $P$ from Tables 2 and 5, where $m_{pl}$ is the mass of the planet). 
This larger mass fraction would result in 
relatively large ancient oceans on Mars and Venus. On the other hand, 
there is the deuterium/hydrogen paradox of Earth's oceans, as the D/H 
ratio is different for oceans and comets.  Pavlov et al. (1999) 
suggested that solar wind-implanted hydrogen on interplanetary dust 
particles could provide the necessary low-D/H component of Earth's 
water inventory.

Our estimate of the migration of water to the early Earth is in 
accordance with Chyba (1989), but
is greater than those of Morbidelli et al. (2000) and Levison et al. (2001).
The latter obtained smaller
values of $M_{ice}$, and we suspect that this is because they did not 
take into account the migration of bodies into orbits with $Q$$<$4.2 
AU and $q$$<$1 AU.
Perhaps this was
because they modeled a relatively small number of objects, and 
Levison et al. (2001) did not take into
account the influence of the terrestrial planets. In our runs the 
probability of a collision of a single object
with a terrestrial planet could be much greater than the total 
probability of thousands of other objects, so the statistics are 
dominated by rare occurrences that might not appear in smaller 
simulations.
The mean probabilities of collisions can  differ by orders of 
magnitude for different JCOs.
Other scientists considered other initial objects and smaller numbers 
of JCOs, and did not find decoupling from Jupiter, which is a rare event.
We believe there is no contradiction between our present results and 
the smaller
migration of former JCOs to the near-Earth space that was obtained in 
earlier work,
including our own papers (e.g. Ipatov and Hahn, 1999), where we used 
the same integration package.

From measured albedos, Fernandez et al. (2001) 
concluded that
the fraction of extinct comets among NEOs and unusual asteroids is significant
(at least 9\% are candidates).
Rickman et al. (2001) believed that comets
played an important and perhaps even dominant role among all km-size 
Earth impactors.
In their opinion, dark spectral classes that might include the 
ex-comets are severely underrepresented
(see also Jewitt and Fernandez, 2001).
Our runs showed that if one observes
former comets in NEO orbits, then it is  probable that they have already
moved in such orbits for millions (or at least hundreds of thousands) 
years, and only a few of them have been
in such orbits for short times (a few thousand years).
Some former comets that have moved in typical NEO orbits for millions 
or even hundreds of millions of years,
and might have had
multiple close encounters with the Sun,
could have lost their mantles,
which causes their low albedo, and so change their albedo (for most
observed NEOs, the albedo is greater than that for comets; Fernandez et al., 2001)
and would look like typical asteroids.
Typical comets have larger rotation periods than typical NEOs 
(Binzel et al., 1992),
but, while losing considerable portions of their masses, extinct 
comets can decrease these periods. In future we plan to consider a 
larger initial number of objects initially located beyond Jupiter
in order to better estimate their
probabilities of migration to a near-Earth space.

\section*{CONCLUSIONS}

Collision statistics for the terrestrial planets
are dominated by very small numbers of bodies 
that reach orbits with high collision probabilities, so it is 
essential to consider very large numbers of particles. The initial 
conditions for the orbit integrations appear to matter more than the 
choice of orbit integrator.  Some Jupiter-family comets can reach 
typical NEO orbits and remain there for millions of years. While the 
probability of such events is small
(about 0.1\% in our runs and perhaps smaller for other initial data), 
nevertheless the majority of collisions of former JCOs with the 
terrestrial planets are due to such objects.  The amount of water 
delivered to the Earth during planet formation could be about the 
mass of the Earth oceans. From the dynamical point of view there 
could be (not 'must be') many extinct comets among the NEOs. For better 
estimates of the portion of extinct comets among NEOs
 we will need orbit
integrations for many more TNOs and JCOs, and wider analysis of observations and craters.  

\section*{ACKNOWLEDGMENTS}

      This work was supported by NRC (0158730), NASA (NAG5-10776), 
INTAS (00-240), and RFBR (01-02-17540).
      For preparing some data for figures we used some subroutines 
written by P. Taylor.
We are thankful to W. F. Bottke, S. Chesley, and H. F. Levison for 
helpful remarks.

\centerline{}

\end{document}